\documentclass[10pt,eqsecnum,twocolumn,pra]{revtex4}
\usepackage{graphicx} 
\usepackage{amsthm}
\usepackage{amsmath}
\usepackage{amssymb}
\usepackage{bbm,amsfonts}
\usepackage{setspace}
\usepackage{nomencl}
\usepackage{amstext} 
\usepackage{subfigure}
\usepackage{color} 
\usepackage{comment} 

\newcommand{\tr}{{\mathrm{Tr}}}

\newcommand{\Wal}{w}
\newcommand{\CPMG}{\text{CPMG}}
\newcommand{\PDD}{\text{PDD}}

\newcommand{\Tmin}{T_{\text{min}}}

\newcommand{\Fish}{\mathcal{I}_f}
\newcommand{\Fishq}{\mathcal{I}_{f,q}}
\newcommand{\etan}{\eta_{\text{0}}}

\newtheorem{theorem}{Theorem}

\newtheorem{definition}{Definition}

\newtheorem{corollary}{Corollary}

\theoremstyle{definition}

\def\>{\rangle}
\def\<{\langle}

\begin{document}

\title{Reconstructing the Profile of Time-Varying Magnetic Fields With Quantum Sensors}

\author{Easwar Magesan, Alexandre Cooper, Honam Yum, Paola Cappellaro}

\affiliation{Department of Nuclear Science and Engineering and Research Laboratory of Electronics,\\Massachusetts Institute of Technology,  Cambridge, Massachusetts 02139, USA}

\begin{abstract}
Quantum systems have shown great promise for  precision metrology thanks to advances in their control. This has allowed not only the sensitive estimation of external parameters but also the reconstruction of their temporal profile. In particular,   quantum control techniques and orthogonal function theory have been applied to the reconstruction of the complete profiles of time-varying magnetic fields. Here, we provide a detailed theoretical analysis of the reconstruction method based on the Walsh functions, highlighting the relationship between the orthonormal Walsh basis, sensitivity of field reconstructions, data compression techniques, and dynamical decoupling theory.  Specifically, we show how properties of the Walsh basis and a detailed sensitivity analysis  of the  reconstruction protocol provide a method to characterize the error between the reconstructed and true fields. In addition, we prove various results about the negligibility function on binary sequences which lead to data compression techniques in the Walsh basis and a more resource-efficient reconstruction protocol. The negligibility proves a fruitful concept to unify  the information content of Walsh functions and their dynamical decoupling power, which makes the reconstruction method robust against noise. 
\end{abstract}

\maketitle

\section{Introduction}


The goal of parameter estimation theory is to use empirical data to estimate a set of unknown parameters of interest. Typically, the unknown parameters describe properties of a physical system that can be measured in some well-defined manner. The measurement results create the set of data from which statistical estimators of the unknown parameters can be obtained.
Quantum parameter estimation (QPE) theory allows both the physical system and sensing procedures to be quantum-mechanical in nature. Many important concepts in classical parameter estimation theory, such as the Fisher information~\cite{Fisher25} and Cram{\'e}r-Rao bound~\cite{Cramer46,Rao45}, have been developed within the framework of QPE~\cite{Helstrom67,BC94}. 

A special focus of QPE is to find how one can (optimally) estimate parameters that determine the state or dynamics of a quantum system~\cite{Hol94,VDM,GW01}. For instance, one may be interested in estimating parameters associated to an underlying Hamiltonian~\cite{CSG}, unitary operation~\cite{Kitaev95}, or general quantum channels~\cite{BPP}. 
Quantum sensors can provide significant improvement over sensors based on classical physics in terms of sensitivity~\cite{GLM} and both spatial ~\cite{GDT} and field amplitude resolution~\cite{JLS}. Quantum phase estimation~\cite{Kitaev95} has been applied to many  practical problems, for example clock synchronization \cite{Jozsa00,Revzen03}, reference frame alignment \cite{Chiribella04}, frequency measurements \cite{Rosenband08}, position measurements \cite{GLM} and magnetic and electric field sensing~\cite{TCC,Dolde11}. Although the methods presented in this paper can, in principle, be applied to all these techniques, for concreteness we will focus on magnetic field sensing with spin qubits.

In spin-based sensing, the magnetic field strength is recorded in the relative phase accumulated by the system when prepared in a state equal to the uniform superposition of its energy eigenstates~\cite{Ram50}. Hence, one can obtain estimates of the field by obtaining information about the relative phase accumulated between the eigenstates. A significant problem in magnetometry is to not just estimate a single parameter quantifying magnetic field strength, but to \emph{estimate (reconstruct) the complete profile of the field over some acquisition period}. Reconstructing the complete field profile can be important in a variety of different physical settings, for instance, in understanding the dynamical behavior of neuronal activity in biological systems. 

Recently, a technique was proposed for experimentally reconstructing the complete profile of time-varying magnetic fields~\cite{CMYC} using spin-based magnetic field sensing techniques. This method allows one to reconstruct magnetic fields with quantifiable analyses of reconstruction error, sensitivity, convergence, and resources. The method was implemented using the single electronic spin in an NV center for a variety of different field profiles and high-fidelity reconstructions were obtained in all cases. In particular, Ref.~\cite{CMYC} experimentally demonstrated the reconstruction of the magnetic field produced by a physical model of a single neuronal action potential. The theory behind the method relies on a variety of topics within classical and quantum information theory, for instance, quantum control theory, QPE, and the use of orthogonal functions in signal processing through the existence of the Walsh basis~\cite{Walsh23}.

This paper serves as a detailed analysis of the theory underlying Ref.~\cite{CMYC} and also provides results that can be of independent interest in signal processing, QPE, magnetometry, and quantum control theory. 
The material of the paper is organized as follows. In Sec.~\ref{sec:setup} we present the main problem we are interested in; how to reconstruct the profile of general deterministic magnetic fields. We also discuss why a digital orthonormal basis of functions, such as the Walsh basis, is an ideal choice to represent the field when performing quantum magnetometry. In Sec.~\ref{sec:Walsh} we provide a detailed review of the digital orthonormal Walsh basis. This section is intended as a reference for properties of the Walsh basis used throughout the rest of the paper. We provide various definitions related to generating and ordering Walsh functions, and discuss the CPMG and PDD~\cite{CP,MG} Walsh functions which have particular importance in quantum information theory. 

Sec.~\ref{sec:Sensitivity} contains a statistical analysis of the Walsh reconstruction method. We calculate the Fisher information and sensitivity in estimating fluctuations of the field via the Walsh reconstruction method and analyze how errors in estimating Walsh coefficients propagate to the time-domain. Finally, we put many of these results together to characterize how close the reconstructed field is to the actual field. The key result is that the reconstructed field is modeled as a random variable whose variance depends on both the sensitivity and error in estimating Walsh coefficients. 

Ideally, one wants to spend  resources to only estimate Walsh coefficients that have significant contribution to the overall reconstruction of the magnetic field. Sec.~\ref{sec:Data Compression} provides various methods for compressing the information contained in the Walsh spectrum of a function. We give an introduction to the concept of negligibility of Walsh coefficients which is used throughout the presentation. We discuss the negligibility of the CPMG and PDD Walsh coefficients and provide a detailed explanation of the structure of the negligibility function. We discuss three methods for performing data compression in the Walsh basis, a CPMG/PDD-based method, threshold sampling, and the sub-degree method~\cite{C-K75}. 

Lastly, noise effects play an important role in real experiments and so it is important to understand how the Walsh reconstruction method performs in the presence of noise. Interestingly, the same control sequences we apply to reconstruct the magnetic field also act as ``dynamical decoupling" (DD) sequences~\cite{VKL} that help preserve the coherence of the quantum sensor. The utility of Walsh sequences for dynamical decoupling techniques has recently been discussed in Ref.~\cite{HKV}. We build on this discussion in Sec.~\ref{sec:Dynamical Decoupling} and use the negligibility function to provide a unifying duality between using the Walsh functions to reconstruct time-varying fields and perform dynamical decoupling.

\section{Reconstructing Temporal Profiles} \label{sec:setup}

We are interested in reconstructing the complete profile of a time-varying field $b(t)$ over some time interval $t\in[0,T]$, where we call $T$ the acquisition time. We perform measurements with a single qubit sensor, in direct analogy to a previous experimental demonstration~\cite{CMYC} using a nitrogen-vacancy (NV) center in diamond. 
Assuming the magnetic field is directed along the quantization $z$-axis of the qubit sensor, the Hamiltonian $\hat{H}(t)$ of the system 
is given by
\begin{equation}
\hat{H}(t)=\left[\frac{\omega_0}{2}+\gamma b(t)\right]\sigma_Z,
\end{equation}
where $\omega_0$ is the frequency splitting of the qubit and $\gamma$ is the coupling strength to the field. For example, in the case of a spin coupled to a magnetic field, $\gamma$ is the gyromagnetic ratio,  e.g., $\gamma_e=28~\text{Hz/nT}$ for NV electronic spin qubits. Moving to the rotating frame, the evolution of the system is given by,
\begin{align}
\hat{U}(t)&=e^{-i\left[\gamma\int_0^Tb(t)dt\right] \sigma_Z}=e^{-i\phi(T)\sigma_Z} .
\end{align}
If $b(t)$ is known to be a constant field, $b(t)=b$, the field amplitude $b$ can be estimated via Ramsey interferometry measurements~\cite{Ram50} by applying a $\frac{\pi}{2}$ pulse about $X$, waiting for a time $T$, implementing a $-\frac{\pi}{2}$ pulse about $Y$, and performing a projective measurement in the computational basis, while repeating the experiment many times to gather measurement statistics.

The probability of obtaining outcome ``0"  (by measuring the qubit in the $|{0}\rangle$ state) is
\begin{align}
p_0&=\frac{1+\sin\left(\gamma\int_{t=0}^Tb dt\right)}{2} = \frac{1+\sin(\gamma bT)}{2},
\end{align}
with the probability of obtaining outcome ``1" given by $p_1=1-p_0$. Hence, under the assumption that $\gamma bT \in \left[-\frac{\pi}{2},\frac{\pi}{2}\right]$, one can determine $b$ from either of the above equations. Note that the range $\left[-\frac{\pi}{2},\frac{\pi}{2}\right]$ sets a limit on the magnitude of $b$ which can be measured, that is, the dynamic range,  $\pi/(\gamma T)$ . 

We now generalize the above discussion to time-varying fields with arbitrary profile in time. Suppose $b(t)$ is given by a deterministic, smooth 
function on the time interval $[0,T]$. The main question is whether there exists a well-defined, systematic procedure to accurately reconstruct $b(t)$ which allows for the development of error, sensitivity, convergence, and resource analyses. In Ref.~\cite{CMYC} we presented a solution to this problem and, as previously mentioned, one of the main goals of this paper is to provide a more detailed description of this method.

\begin{figure}[t]
\begin{center}
\includegraphics[width=0.45\textwidth]{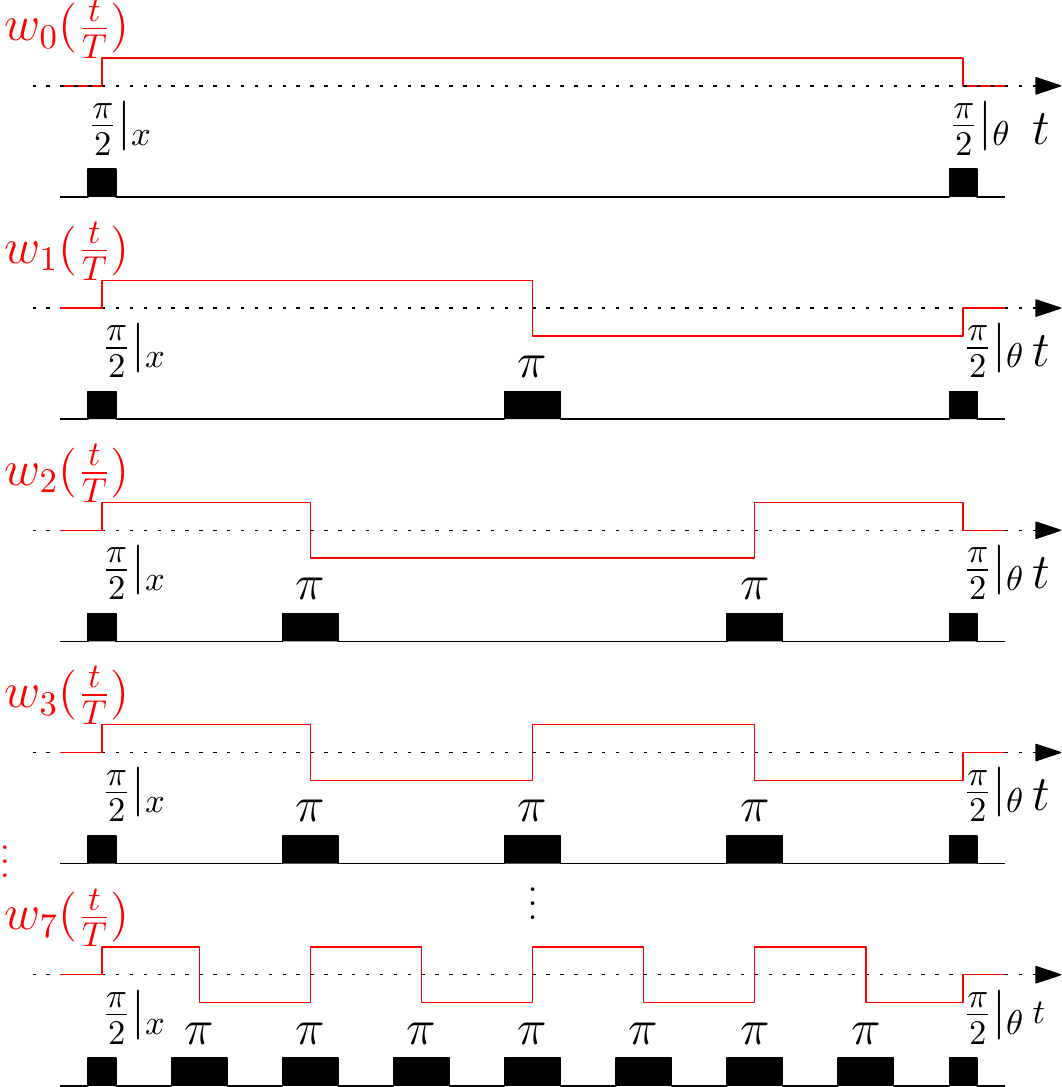}
\caption{\label{Fig:walsh_protocol} Control sequences derived from Walsh functions (in sequency ordering). The first sequence (Ramsey) describes the simplest parameter estimation protocol with no modulating $\pi$ pulses. The second sequence corresponds to the Hahn spin echo and, in general, the $m$'th Walsh function requires $m$ $\pi$ pulses in the control sequence.}
\end{center}
\end{figure}

The key is to note that during the acquisition period $T$, one has the opportunity to modulate the evolution of the quantum system with control sequences of sign-inverting $\pi$-pulses (see Fig.~\ref{Fig:walsh_protocol}). When $b(t)$ is of a simple known form, standard control sequences have been proposed to perform reconstruction of $b(t)$~\cite{TCC,MSH}. In particular, when $b(t)$ is known to be a sinusoidal function with known frequency $\nu$ and phase $\alpha$,
\begin{align}
b(t)&=b\sin(2\pi\nu t + \alpha),
\end{align}
implementing a Hahn spin-echo sequence~\cite{Hahn50} with a single $\pi$-pulse at the zero-crossing of the sine function, which corresponds to the midpoint of the evolution period, allows one to estimate the amplitude of the field $b$. Our goal is to make this scheme as general as possible in that we want to reconstruct the entire temporal profile of an \emph{unknown} time-varying field $b(t)$. We emphasize that $b(t)$ does not need to be a trigonometric function, or even known \emph{a priori}.

Our solution, presented in Ref.~\cite{CMYC}, is as follows. During the acquisition period $T$, one can apply specific \emph{control sequences} that consists of $\pi$ rotations about the $X$ (or $Y$) axis at pre-determined times $t_j$ in $[0,T]$. The key is that these control sequences can be chosen to map in a one-to-one manner to a known piecewise constant orthonormal basis of the set of square-integrable functions, $L^2[0,T]$, on $[0,T]$. One example of such a basis is the set of orthonormal Walsh functions, which we discuss in detail in Sec.~\ref{sec:Walsh}. Let us outline our protocol in complete generality for when $b(t)$ is unknown and the orthonormal basis is left arbitrary. 

Suppose we partition $[0,T]$ into $n$ uniformly spaced intervals with endpoints $t_j=\frac{jT}{n}$ for $j \in \{0,...,n\}$. At each $t_j$, for $j \in \{0,...,n-1\}$, a $\pi$-pulse is applied according to some pre-defined $n$ length bit string $\alpha$. The occurrence of a $1$ in $\alpha$ indicates that a $\pi$-pulse should be applied and a $0$ indicates that no $\pi$-pulse should be applied. Under this binary encoding of pulse sequences, the evolution of the system is given by
\begin{align}
U(T)&=\: \Pi_{j=n-1}^0\left\{\left[e^{-i\left[\gamma \int_{t_j}^{t_{j+1}}b(t)dt\right]\sigma_Z}\right]X_\pi^{\alpha(j)}\right\} \\\nonumber&=e^{-i\left[\gamma\int_0^T\kappa_\alpha(t) b(t)dt\right] \sigma_Z} = e^{-i \phi_\alpha(T) \sigma_Z}
\end{align}
where $X_\pi$ corresponds to a $\pi$-rotation about the $X$ axis in the Bloch sphere representation.
It is straightforward to verify that $\kappa_\alpha(t)$ is a piecewise constant function taking values $\pm 1$ on each $(t_j,t_{j+1})$ and a switch $1 \leftrightarrow -1$ occurs at $t_j$ if and only if a $\pi$-pulse is implemented at $t_j$. Hence, performing a $\pi$-modulated Ramsey experiment produces a phase evolution $\phi_\alpha(T)$ given by the scaled inner product between $\kappa_\alpha(t)$ and $b(t)$,
\begin{equation}
\phi_\alpha(T)\!=\!\gamma T \langle \kappa_\alpha(t),b(t)\rangle\!=\!\gamma T \left[\frac{1}{T}\!\int_0^T\!\kappa_\alpha(t) b(t)dt\right]\!.
\end{equation}
Upon rotation by a $-\frac{\pi}{2}$ pulse about $Y$, a measurement in the computational basis yields the probability 
\begin{align}
p_0&=\frac{1+\sin\left(\phi_\alpha(T)\right)}{2} = \frac{1+\sin(\gamma T \langle \kappa_\alpha(t),b(t)\rangle)}{2}
\label{eq:p0}
\end{align}
and $p_1=1-p_0$. Hence, if 
$
\gamma T \left|\langle \kappa_\alpha(t),b(t)\rangle \right| \leq \frac{\pi}{2}
$,  
one can estimate $\langle \kappa_\alpha(t),b(t)\rangle$ exactly from $p_0$. Now, suppose the set of all $\kappa_\alpha (t)$ forms an orthonormal basis of $L^2[0,T]$. In this case, we can write any $b(t)$ as
\begin{align}
b(t)&=\sum_\alpha \langle \kappa_\alpha(t),b(t)\rangle \kappa_\alpha(t),
\end{align}
and so being able to estimate all $\langle \kappa_\alpha(t),b(t)\rangle$ implies one can reconstruct $b(t)$ exactly. As mentioned previously, one can take the Walsh basis $\{w_m(t)\}$ to represent the different $\kappa_\alpha(t)$. Since the Walsh basis is countably infinite, it can be ordered from $m=0$ to $\infty$, so that
\begin{align}
b(t)&=\sum_{m=0}^\infty \langle w_m(t),b(t)\rangle w_m(t).
\end{align}

In theory, this protocol will reconstruct $b(t)$ exactly.  In a practical implementation, some questions arise that we will address in the reminder of the paper:

\begin{enumerate}
\item What is the sensitivity in estimating each of the $\langle w_m(t),b(t)\rangle$ and how does this affect the overall reconstruction? (see Sec.~\ref{sec:Sensitivity})
\item How does the reconstruction behave if we truncate the decomposition series to a finite order at a certain point? Are there certain coefficients that we can neglect?  (Sec.~\ref{sec:Data Compression})
\item How does the control sequence defined by $w_m$ aid in preserving the coherence of the system? (Sec.~\ref{sec:Dynamical Decoupling})
\end{enumerate}
Since the Walsh functions form an orthonormal basis that is well adapted to the reconstruction of time-varying fields with digital filters, we briefly review their properties, focusing on the ones that are the most relevant to the reconstruction method.

\section{The Walsh Basis} \label{sec:Walsh}

\subsection{Walsh Functions, Ordering, and Partial Reconstructions}\label{sec:Definition}
Orthogonal functions play a prominent role in a vast array of theoretical and applied sciences, for instance, in communication theory, mathematical modeling, numerical analysis, and virtually any field that utilizes signal processing theory. 
An important problem is to use the physical nature of the signal, control, and sensing mechanisms to determine what mathematical basis is best suited for processing the physical information. The sine and cosine trigonometric functions are the most commonly used set of orthogonal functions for a variety of reasons; they have clear physical interpretation, model a large set of physical phenomena, and are easy to describe mathematically. Moreover, the advent of the fast Fourier transform has made it more computationally efficient to move between the time and frequency domains. In analog signal processing, these functions are often the method of choice for decomposing and manipulating signals. 

Non-trigonometric sets of orthogonal functions have also found great utility in many different scientific fields~\cite{Davis}. For instance, orthogonal polynomial bases such as Chebyshev and Hermite polynomials are used extensively in numerical analysis, the Haar basis is used in pattern recognition and Bessel functions are used for solving problems related to heat conduction and wave propagation. In digital communication theory, the \emph{Walsh basis}~\cite{Walsh23} has found a great deal of success due to the natural ``digital" (piecewise constant) behavior of its constituent functions. The Walsh functions are useful in digital signal processing because they only take the values $\pm 1$, yet also constitute a complete orthonormal basis of $L^2[0,1]$.  
Moreover, unlike the Fourier transform which requires a large set of multiplication operations, the Walsh transform, which moves between the time and ``sequency" domains (defined below), only requires a straightforward set of addition operations. We now discuss the Walsh basis in detail.

\begin{table*}
\begin{tabular}{| c | c | c | c | c | c}
  \cline{1-2} \cline{4-5}
  \multicolumn{2}{|c|}{Paley} & \multicolumn{1}{c}{~} & \multicolumn{2}{|c|}{Sequency} \\
  \hline       
  $w_m$ & Binary & $m$ & Gray & $w_{m}$ \\
  \hline                        
  $R_0R_0R_0$ & 000 & 0 & 000 & $R_0R_0R_0$\\
  $R_0R_0R_1$ & 001 & 1 & 001 & $R_0R_0R_1$\\
  $R_0R_2R_0$ & 010 & 2 & 011 & $R_0R_2R_1$\\
  $R_0R_2R_1$ & 011 & 3 & 010 & $R_0R_2R_0$\\
  \vdots & \vdots & \vdots & \vdots & \vdots \\
  $\displaystyle\Pi_{k=1}^{n}\left[R_k(t)\right]^{m_k}$ & $m_n\cdots m_2m_1$ & $2^n-1$ & $g_n\cdots g_2g_1$ & $\displaystyle{\Pi_{k=1}^{n}}\left[R_k(t)\right]^{g_k}$\\
  \hline  
\end{tabular}
\caption{\label{tab:sequences}Paley and sequency ordering of Walsh functions}
\end{table*}

For each natural number $m$ we have in standard binary ordering,
\begin{align}
m=&\sum_{k=1}^{n}\frac{m_k}{2^k}= \frac{m_1}{2^1}+\frac{m_2}{2^2}+\cdots+\frac{m_n}{2^{n}}\\&\rightarrow 0.m_1m_2\cdots m_n \rightarrow (m_n,...,m_2,m_1).\nonumber
\end{align}
Now, the $m$'th Walsh function in \emph{Paley ordering}~\cite{Paley32} is given by 
\begin{align}
w_m(t)&=\displaystyle{\Pi_{k=1}^{n}}\left[R_k(t)\right]^{m_k}= \displaystyle{\Pi_{k=1}^{n}}(-1)^{m_kt_k},
\end{align}
where $R_k:[0,1]\rightarrow \mathbb{R}$ is the $k$'th Rademacher function~\cite{Rad22}, $R_k(t)=(-1)^{t_k}$ (see Table~\ref{tab:sequences}).
Note that the $k$'th Rademacher function $R_k(t)$ is just a periodic square wave function on [0,1] that oscillates between $+1$ and $-1$ and has $2^k$ intervals with $2^k-1$ jump discontinuities. More precisely, $R_0(t)=1$ for all $t$, and if $k\geq 1$, the $k$'th Rademacher function is defined as 
\begin{align}
R_k(t) &= \left\{\begin{array}{ll} 1 & : t \in [0, 1/2^k) \\ -1 & : t \in [1/2^k, 1/2^{k-1}) \end{array} \right.,
\end{align}
 extended periodically to the unit interval $[0,1]$. Another alternative representation of the $k$'th Rademacher function ($k\geq 0$) is 
 \begin{equation}
R_k(t)=\text{sgn}\left(\sin(2^k\pi t)\right).
\end{equation}

The Walsh functions in \emph{sequency ordering} are obtained by multiplying the Rademacher functions according to the Gray code~\cite{Gray53} of the binary sequence of $m$ (see Table~\ref{tab:sequences}).
Gray code is the ordering of binary sequences where two successive sequences can only differ in exactly one digit, and right-most digits vary fastest. If the integer $m$ is written in binary expansion with respect to the Gray code, $g=\text{Gray}(m)=g_n\cdots g_2g_1$, then the $m$'th Walsh function is given by
\begin{align}
w_m(t)&=\displaystyle{\Pi_{k=1}^{\infty}}\left[R_k(t)\right]^{g_k}= \displaystyle{\Pi_{k=1}^{\infty}}(-1)^{g_kt_k}.
\end{align}

Since the sequency and Paley orderings differ only in terms of how each integer $m$ is represented in terms of binary sequences, there exists a linear transformation for going from one ordering to the other, and the transformation reduces to switching between the Gray code and binary code ordering.  Note also that the set of the first $2^k$ sequency-ordered Walsh functions is equal to the set consisting of the first $2^k$ Paley ordered Walsh functions. 


There are other useful orderings of the Walsh functions~\cite{ZQ}, however for the remainder of this paper we will focus on Walsh functions in Paley ordering and sequency ordering. There is no general guideline that governs when each type of ordering should be used; however, it is common that in mathematical analysis of Walsh functions, Paley ordering is used, while in signal processing, engineering, and communication theory applications, sequency ordering is used. 

Since the Walsh functions form an orthonormal basis, each square-integrable function $f \in L^2[0,T]$ 
can be written as
\begin{eqnarray}
f(t)&=& \sum_{m=0}^{\infty}\hat{f}_m w_m\left(\frac{t}{T}\right),
\end{eqnarray}
where $\hat{f}_m$ is the $m$'th \emph{Walsh coefficient} of $f(t)$,
\begin{eqnarray}
\hat{f}_m&=& \frac{1}{T}\int_0^Tf(t)w_m\left(t/T\right)dt \nonumber,
\end{eqnarray}
and $w_m:[0,1]\rightarrow \mathbbm{R}$ 
 is the $m$'th \emph{Walsh function}.

The $N$'th partial sum of $f(t)$, $f_N$, is defined as the truncation of the Walsh series of $f(t)$ to its first $N$ terms,
\begin{eqnarray}
f_N(t)&=& \sum_{j=0}^{N-1}\hat{f}_j\Wal_j\left(\frac{t}{T}\right).
\end{eqnarray}
It can be shown that if $f$ is continuous then, as $N\rightarrow \infty$, $f_N(t)$ converges uniformly to $f(t)$. We also define the $n$'th order reconstruction of $f\in L^2[0,T]$ to be the $2^n$'th partial sum of $f$. We now discuss some important subsets of Walsh functions in quantum information theory, in particular, the CPMG and PDD functions.

\subsection{CPMG and PDD Walsh Functions}\label{sec:CPMGandPDD}

The ``Carr-Purcell-Meiboom-Gill (CPMG)" pulse sequences~\cite{CP,MG} originated in NMR as an active control method to extend the coherence of quantum systems. They are natural extensions of the  Hahn spin echo sequence~\cite{Hahn50}. Recently, CPMG sequences have also been used to reconstruct the noise power spectrum of quantum environments~\cite{CLL,ASG,BGY,AS,BPB}. The CPMG sequences can be associated to the set of ``CPMG Walsh functions" by associating $\pi$-pulses to the points of discontinuity in each CPMG Walsh function. The CPMG Walsh functions  are the $\Wal_j$ with  $j\in\{2^k\}_{k=1}^\infty$ in sequency ordering 
and   $j\in\{\frac32 2^k\}_{k=1}^\infty$ in Paley ordering.
We define the $M$'th CPMG partial sum of $f$ to be 
\begin{align}
f^{CPMG}_M(t)&=\sum_{j=1}^{M}\hat{f}_{2^j}\Wal_{2^j}(t)\nonumber \\
\end{align}
where the indexing is with respect to sequency ordering.
%

Similarly to CPMG sequences, the ``periodic dynamical decoupling (PDD)" sequences~\cite{CP} are also useful extensions of the Hahn spin echo sequence. The set of PDD functions corresponds  to the $\Wal_j$ with $j\in\{2^k-1\}_{k=1}^\infty$ in sequency ordering and $j\in\{2^{k-1}\}_{k=1}^\infty$ in Paley ordering. Hence, the PDD functions correspond exactly to the set of Rademacher functions. 
 Note that while every CPMG function  is symmetric about the midpoint of the domain, every PDD function  is anti-symmetric.
We also define the $M$'th PDD partial sum of $f$ to be 
\begin{align}
f^{PDD}_M(t)&= \sum_{j=1}^{M}\hat{f}_{2^j-1}\Wal_{2^j-1}(t)
\end{align}
where again the indexing is with respect to sequency ordering.

Note that a $k$'th order Walsh reconstruction of $f$ contains exactly $k-1$ CPMG and $k$ PDD Walsh functions. Thus, there are only logarithmically many CPMG and PDD Walsh coefficients in a $k$'th order reconstruction of $f$. One can obtain a simple resource analysis between implementing an $M$'th partial sum and implementing an $M$'th CPMG or PDD partial sum of $f$. Indeed, the $M$'th $\CPMG$ or $\PDD$ partial sum of $f$ contains an exponentially large number of zero-crossings (over all functions in the partial sum) compared to only a polynomial number of zero-crossings in an $M$'th partial sum of $f$. It is straightforward to see why there is an exponential gap between the partial sums. First, note that the $i$'th Walsh function contains $i$ crossings (that is $i$ $\pi$ pulses must be applied to implement this particular sequence). Hence, the total number of zero-crossings contained in the $M$'th partial sum is
\begin{equation}
\sum_{i=0}^{M-1}i = \frac{M(M-1)}{2}.
\end{equation}
On the other hand, the $M$'th CPMG partial sum consists of 
\begin{align}
 \sum_{j=1}^{M-1}2^j&=  2^M-2
\end{align}
total zero-crossings and the $M$'th PDD partial sum consists of
\begin{align}
 \sum_{j=1}^{M}(2^j-1)&= 2^{M+1}-(M+2)
\end{align} 
total zero-crossings. Thus, each of the $M$'th CPMG and PDD partial sums contain an exponential number of crossings while the $M$'th Walsh partial sum contains only a polynomial number of crossings.

\section{Sensitivity and Reconstruction Error}\label{sec:Sensitivity}
In this section, we analyze the performance of  the Walsh reconstruction method in terms of its sensitivity, error propagation, and reconstruction error. First, we calculate the sensitivity in estimating magnetic field amplitudes with Walsh coefficients. Next, we obtain an expression for the sensitivity in estimating individual Walsh coefficients and then analyze how errors in the estimated Walsh coefficients propagate into the time-domain reconstruction. Finally, we combine these results with known bounds on truncation of Walsh series to give an explicit expression for how much the reconstructed field can deviate from the true field.

%
%
%
%

\subsection{Sensitivity in Estimating Magnetic Field Amplitudes}\label{sec:sensamp}
Suppose  $b(t)=bf(t)$ for some square-integrable function $f(t)$. The sensitivity, $\eta$, in estimating the amplitude $b$ from a classical parameter estimation scheme is given by
\begin{align}\label{eq:etadef}
\eta&=\frac{1}{\sqrt{\Fish(b)}},
\end{align}
where $\Fish(b)$ is the Fisher information~\cite{Fisher25} with respect to the unknown parameter $b$. The Fisher information is defined in terms of the likelihood function of $b$, which is a measure of how likely it is that the parameter we want to estimate (here being $b$) is equal to a particular value (given a set of data collected by performing measurements on the system). The Fisher information can be thought of as a measure of the local curvature of the likelihood function around $b$. If the likelihood function is very flat, then little information is gained on average from sampling (and the Fisher information is small), while if the likelihood function is sharply peaked around $b$, a large amount of information can be gained on average from sampling (and the Fisher information is large).  

Formally, if $X$ is a classical random variable that we sample from ($X$ contains information about $b$), and we let $L(X;b)$ denote the likelihood function, then the Fisher information is defined as
\begin{align}\label{eq:FisherInformation}
\Fish&=\mathbbm{E}\left[\left(\frac{\partial}{\partial b}\log L(X;b)\right)^2 \Bigg| \: b\right],
\end{align}
where the expectation value is taken with $b$ fixed. The Cram{\'e}r-Rao bound~\cite{Cramer46} states that the variance of \emph{any} unbiased estimator of $b$, denoted $\tilde{b}$, is fundamentally limited (bounded below) by
\begin{align}
\text{Var}\left(\tilde{b}\right)\geq \frac{1}{\Fish(b)}.
\end{align}
Hence, the Cram{\'e}r-Rao bound and Eq.~(\ref{eq:etadef}) imply that the sensitivity $\eta$ is equal to the minimum possible standard deviation of any unbiased estimator $\tilde{b}$ of $b$. 

Extensions of these concepts to the quantum information setting have been given~\cite{Helstrom67,BC94} where quantum states $\rho_b$ now represent the (non-commutative) random variables we sample from. For each possible quantum measurement (POVM) $\{E_\beta\}$, one can define a measurement-dependent Fisher information according to Eq.~(\ref{eq:FisherInformation}) and the Born rule 
\begin{align}
\text{pr}(\beta)&=\tr(\rho_b E_\beta),
\end{align}
which is the probability of obtaining outcome ``$\beta$". Optimizing over all possible POVM's leads to the \emph{quantum Fisher information}
\begin{align}
\Fishq (b)&=\tr\left(\rho_bL_b^2\right) =\tr\left(L_b\frac{\partial\rho_b}{\partial b}\right),
\end{align}
where $L_b$ is the self-adjoint solution to the symmetric differential equation
\begin{align}
\frac{\partial\rho_b}{\partial b}&=\frac{\rho_bL_b+L_b\rho_b}{2}.\label{eq:symmetricderivative}
\end{align}
This gives the \emph{quantum Cramer-Rao bound}~\cite{BC94,Helstrom67}
\begin{align}
\text{Var}\left(\tilde{b}\right)\geq \frac{1}{\Fishq(b)} = \frac{1}{\tr\left(L_b\frac{\partial\rho_b}{\partial b}\right)},
\end{align}
where $\tilde{b}$ is now any unbiased estimator that results from allowing the system and measurement to be quantum-mechanical. Hence, from Eq.~(\ref{eq:etadef}), the sensitivity is given by
\begin{align}
\eta&=\frac{1}{\sqrt{\tr\left(L_b\frac{\partial\rho_b}{\partial b}\right)}}.
\end{align}
In practice, one repeats the sensing procedure $M$ times to enhance the sensitivity by
\begin{align}
\eta&=\frac{1}{\sqrt{M\tr\left(L_b\frac{\partial\rho_b}{\partial b}\right)}}.
\end{align}
To have a standardized measure of the sensitivity, one typically defines the normalized version
\begin{align}
\etan&=\sqrt{M}\eta.
\end{align}


Suppose our goal is to estimate the $\alpha$'th Walsh coefficient of $b(t)$. Assuming the system is immune to decoherence effects, the state after the initial $\frac{\pi}{2}$ pulse and modulating Walsh sequence is given by
\begin{align}
\rho_b&=
\frac{1}{2}\left( \begin{array}{ccc}
1 & e^{-i\phi_\alpha} \\
e^{i\phi_\alpha} & 1 \end{array} \right),
\end{align}
where 
\begin{align}\label{eq:phi_alpha}
\phi_\alpha&=\gamma b T\hat{f}_\alpha.
\end{align}
Thus
\begin{align}
\frac{\partial \rho_b}{\partial b}&=
\frac{1}{2}\left( \begin{array}{ccc}
0 & \frac{-i\phi_\alpha}{b}e^{-i\phi_\alpha} \\
\frac{i\phi_\alpha}{b}e^{i\phi_\alpha} & 0 \end{array} \right),
\end{align}
and so Eq.~(\ref{eq:symmetricderivative}) can be written as
\begin{align}
\left( \begin{array}{ccc}
0 & \frac{-i\phi_\alpha}{b}e^{-i\phi_\alpha} \\
\frac{i\phi_\alpha}{b}e^{i\phi_\alpha} & 0 \end{array} \right)
&=L_b\left( \begin{array}{ccc}
1 & e^{-i\phi_\alpha} \\
e^{i\phi_\alpha} & 1 \end{array} \right)\nonumber \\
&\: \: \: \: \: +
\left( \begin{array}{ccc}
1 & e^{-i\phi_\alpha} \\
e^{i\phi_\alpha} & 1 \end{array} \right)L_b.\label{eq:symmetriccoords}
\end{align}
Since $L_b$ is self-adjoint let us suppose
\begin{align}
L_b&=
\left( \begin{array}{ccc}
a_1 & c \\
c^* & a_2 \end{array} \right),
\end{align}
where $a_1$ and $a_2$ are assumed to be real. From Eq.~(\ref{eq:symmetriccoords}) we can obtain
\begin{align}
\frac{i\phi_\alpha}{b}e^{i\phi_\alpha}&=c^*+\frac{(a_1+a_2)}{2}e^{i\phi_\alpha},\\
\frac{-i\phi_\alpha}{b}e^{-i\phi_\alpha}&=c+\frac{(a_1+a_2)}{2}e^{-i\phi_\alpha},
\end{align}
which gives
\begin{align}
\frac{c^*e^{-i\phi_\alpha}-ce^{i\phi_\alpha}}{2}&=\frac{i\phi_\alpha}{b}.\label{eq:finalsymmetric}
\end{align}

Now, note that
\begin{align}
\Fishq&=\tr\left(L_b\frac{\partial\rho_b}{\partial b}\right)\nonumber \\
&=
\frac{1}{2}\tr\left[\left( \begin{array}{ccc}
a_1 & c \\
c^* & a_2 \end{array} \right)
\left( \begin{array}{ccc}
0 & \frac{-i\phi_\alpha}{b}e^{-i\phi_\alpha} \\
\frac{i\phi_\alpha}{b}e^{i\phi_\alpha} & 0 \end{array} \right)
\right]\nonumber \\
&=\frac{i\phi_\alpha}{b}\frac{\left(ce^{i\phi_\alpha}-c^*e^{-i\phi_\alpha}\right)}{2}.\label{eq:traceres}
\end{align}
Hence, by Eq.~(\ref{eq:phi_alpha}), Eq.'s~(\ref{eq:finalsymmetric}) and~(\ref{eq:traceres}) give
\begin{align}
\Fishq&=\frac{\phi_\alpha^2}{b^2}=\gamma^2T^2\hat{f}_\alpha^2,
\end{align}
and so the sensitivity is
\begin{align}
\eta&=\frac{1}{\gamma T\left|\hat{f}_\alpha\right|}.
\end{align}

In practical applications, decoherence and other experimental errors limit the signal visibility, $v_\alpha\leq1$, which depends on the Walsh pulse sequence. Thus the sensitivity is not only set by the Walsh coefficient, but also by the dynamical decoupling power of the corresponding Walsh sequence  (see Section~\ref{sec:Dynamical Decoupling}). 
Suppose the visibility decay is given by $v_\alpha=(e^{-T/T_2(\alpha)})^{p(\alpha)}$, where  $T_2(\alpha)$ and $p(\alpha)$  characterize the decoherence of the qubit sensors during the $\alpha$-th Walsh sequence in the presence of a specific noise environment. In this case we have that $\rho_b$ is given by
\begin{align}
\rho_b&=
\frac{1}{2}\left( \begin{array}{ccc}
1 & v_me^{-i\phi_\alpha} \\
v_me^{i\phi_\alpha} & 1 \end{array} \right).
\end{align}
It is straightforward to verify using the exact same method as above that in this case
\begin{align}
\Fishq&=\gamma^2T^2v_\alpha^2\hat{f}_\alpha^2,
\end{align}
and so
\begin{align}
\eta&=\frac{1}{\gamma Tv_\alpha\left|\hat{f}_\alpha\right|}.\\
\end{align}
If we repeat the measurement procedure $M$ times then
\begin{align}
\Fishq&=M\gamma^2T^2v_\alpha^2\hat{f}_\alpha^2,
\end{align}
and
so
\begin{align}
\eta &= \frac{1}{\sqrt{M}\gamma Tv_\alpha \left|\hat{f}_\alpha\right|}, \nonumber \\
\etan&=\frac{\sqrt{M}}{\sqrt{M}\gamma Tv_\alpha \left|\hat{f}_\alpha\right|} =\frac{1}{\gamma T v_\alpha \left|\hat{f}_\alpha\right|}.\label{eq:sensitivityT2}
\end{align}

\subsection{Sensitivity and Error Propagation}\label{sec:Error}

In practical situations, the estimated Walsh coefficients will have finite errors due to statistical uncertainty from finite sampling, small fluctuations of the field we are measuring, and systematic errors associated with the acquisition protocol (such as imperfect pulses and timing errors). These errors propagate into the time-domain reconstruction. 
One of the advantages of using the Walsh reconstruction method is that, since each Walsh function takes only the values $\pm 1$, a straightforward error propagation can be obtained in many relevant cases of interest. 

Ideally, the reconstructed field has the form
\begin{align}
b_\text{rec}(t)&=\sum_{\alpha \in \mathcal{J}}\hat{b}_\alpha w_\alpha(t)
\end{align}
where $\mathcal{J}$ represents the set of measured Walsh coefficients. However, in the presence of errors, the reconstructed field is a random variable at each $t$,
\begin{align}
B_\text{rec}(t)&=\sum_{\alpha \in \mathcal{J}}\hat{B}_\alpha w_\alpha(t).
\end{align}
Here, errors are introduced only into the estimated Walsh coefficients (the Walsh functions are always exact). Suppose one uses $M$ repetitions to measure each coefficient. For now, let us suppose errors in estimating individual Walsh coefficients are statistical in nature (finite sampling). In the case where $M$ is large, by the central limit theorem, each estimated coefficient is modeled by a Gaussian random variable $\hat{B}_\alpha$ with a normal distribution
$\hat{B}_\alpha\!\sim\!\mathcal{N}\left(\hat{b}_\alpha,\Delta \hat{b}_\alpha^2\right)$.
The notation $\mathcal{N}(\mu,\sigma^2)$ is used to indicate a normal distribution of mean $\mu$ and variance $\sigma^2$. Thus, each $\hat{B}_\alpha$ is normally distributed with mean equal to the ideal coefficient $\hat{b}_\alpha$ and variance $\Delta \hat{b}_\alpha^2$. 

Now, for each $t$, $w_\alpha(t)=\pm1$.
Thus, $B_\text{rec}(t)$ is a linear combination of the normal random variables $\hat{B}_\alpha$ and coefficients $\pm 1$. Hence, for each $t$, $B_\text{rec}(t)$ is a Gaussian random variable with mean equal to the perfectly reconstructed function $b_\text{rec}(t)=\sum_{\alpha \in \mathcal{J}}\hat{b}_\alpha w_\alpha(t)$
and variance $\sum_{\alpha \in \mathcal{J}}\Delta \hat{b}_\alpha^2$. So, \emph{for each $t$}
\begin{align}
B_\text{rec}(t)\sim \mathcal{N}\left(b_\text{rec}(t), \sum_{\alpha \in \mathcal{J}}\Delta \hat{b}_\alpha^2\right).
\end{align}
It is important to note that, since the variance $\sum_{\alpha \in \mathcal{J}}\Delta \hat{b}_\alpha^2$ is time-independent, the errors in estimating Walsh coefficients are evenly spread out in the time-domain representation of $b(t)$. Hence, the time-domain reconstruction is, on average, equal to the exact reconstruction $b_\text{rec}(t)$ with a variance $\sum_{\alpha \in \mathcal{J}}\Delta \hat{b}_\alpha^2$ that is independent of $t$.

We now describe how to model the variance $\Delta \hat{b}_\alpha^2$. Since, for now, we are only considering statistical error, we can use the sensitivity analysis in Subsec.~\ref{sec:sensamp} and set $\Delta \hat{b}_\alpha$ equal to the sensitivity in estimating $\hat{b}_\alpha$. We have that the sensitivity in estimating the $\alpha$'th Walsh coefficient is given by $\frac{1}{\sqrt{M}T\gamma v_\alpha}$ and so
\begin{align}
\Delta \hat{b}_\alpha& = \frac{1}{\sqrt{M}T\gamma v_\alpha}.
\end{align}
Hence
\begin{align}
\sum_{\alpha \in \mathcal{J}}\Delta \hat{b}_\alpha^2&=\frac{1}{MT^2\gamma^2}\sum_{\alpha \in \mathcal{J}}\frac{1}{v_\alpha^2},
\end{align}
which gives
\begin{align}\label{eq:Brecrandom}
B_\text{rec}(t)\sim \mathcal{N}\left(b_\text{rec}(t), \frac{1}{MT^2\gamma^2}\sum_{\alpha \in \mathcal{J}}\frac{1}{v_\alpha^2}\right).
\end{align}
Thus, the reconstructed field has a variance of $\frac{1}{MT^2\gamma^2}\sum_{\alpha \in \mathcal{J}}\frac{1}{v_\alpha^2}$ that forms a time-independent envelope around the exact reconstruction $b_\text{rec}(t)$. The size of the variance depends on the number of repetitions $M$, the acquisition time $T$, and the visibility for each coefficient, $v_{\alpha}$.

In the case of also accounting for experimental errors, such as imperfect pulse and timing errors, there is no method that can account for all possible scenarios. One possible method to account for non-systematic experimental errors is to scale the variance in Eq.~(\ref{eq:Brecrandom}) by some amount so the error envelope surrounding $b_\text{rec}(t)$ is larger. If systematic experimental error is present, one could also place an error on the mean value of $b_\text{rec}(t)$.

\subsection{Putting It All Together: The Reconstruction Error}\label{sec:Reconstruction Error}


Suppose that $B_\text{rec}$ is obtained by truncating the reconstruction at some  order $n$:
\begin{align}
B_\text{rec}(t)&=B_{2^n}(t) \sim \mathcal{N}\left(b_{2^n}(t), \frac{1}{MT^2\gamma^2}\sum_{\alpha=0}^{2^n-1}\frac{1}{v_\alpha^2}\right).
\end{align}
In order to know  how close $B_\text{rec}(t)$ approximates $b(t)$ we need to take into account not only the errors $\Delta \hat{b}_\alpha$, but also  how well the truncation $b_{2^n}(t)$ approximates $b(t)$. The truncation error in $b_{2^n}(t)$ can be bounded by~\cite{C-K75}
\begin{align}
\left|b(t)-b_{2^n}(t)\right|\leq  2^{-(n+1)}T\: \text{max}_{t\in[0,T]}|b^{\prime}(t)|,
\end{align}
where $b^{\prime}(t)$ is the time derivative of $b$.
It is important to emphasize that this bound is not tight since there are situations when the reconstruction is extremely good even though the first derivative diverges. For instance, it is easy to construct a sequence of smooth functions that converges point wise to $w_1(t)$. Hence, while this upper bound provides a sufficient condition for a low-error reconstruction, it is not a necessary condition. 

Thus, for each $t$, the random variable $B(t)$ for the true field is normally distributed with variance
\begin{align}
\frac{1}{MT^2\gamma^2}\sum_{\alpha=0}^{2^{n-1}}\frac{1}{v_\alpha^2}
\end{align}
and mean that can deviate from the reconstructed value $b_{2^n}(t)$ by the amount
\begin{align}
2^{-(n+1)}T\: \text{max}_{t\in[0,T]}|b^{\prime}(t)|.
\end{align}
Again, a possible method of taking experimental error into account is to modify the mean and variance parameters.

In total, given finite resources, one has to find a compromise between applying the available resources to minimize the variance or the reconstruction error. This choice can be informed by results on data compression that we present in the next section.

\section{Data Compression}\label{sec:Data Compression}
In many practical cases, most expansion coefficients in the infinite Walsh series of a function are small or exactly zero, thus reducing the resources needed for a perfect reconstruction. 
Similar results 
are well-known in Fourier theory:  If a function $f$ has $r$ continuous derivatives and $V_{0}^{2\pi}\left[f^{(r)}\right] < \infty$ then, for all $k \neq 0$, the complex Fourier coefficients of $f$ satisfy  
\begin{align}
|c_k|\leq \frac{V_0^{2\pi}\left[f^{(r-1)}\right]}{|k|^r}.
\end{align}
Here, for any $k\geq 1$, $f^{(k)}$ denotes the $k$'th derivative of $f$ and the total variation, $V_{a}^{b}(f)$, of a real-valued, differentiable function $f$ on $[a,b]$ (with $\left|f^{\prime}\right|$ Riemann integrable) is
\begin{equation}
V_{a}^{b}[f]=\int_a^{b} |f^{\prime}(x)| dx.
\end{equation}
Thus, $V_{a}^{b}[f]$ is just the total arc length of the graph of $f(t)$ defined on $[a,b]$.

Similar data compression techniques have been studied for functions represented in the Walsh basis~\cite{C-K72}. The goal of this section is to elucidate and use these techniques to achieve more efficient sampling of Walsh coefficients with finite resources. Ideally, one would like to identify the Walsh coefficients that provide the most information about the field that is being measured. We utilize the \emph{negligibility} function~\cite{C-K72} as it provides information about the negligible coefficients that could be ignored when reconstructing time-varying fields. The negligibility is also useful for understanding the behavior of particular subsets of the Walsh basis, such as the CPMG and PDD functions. 

\subsection{Negligibility and Bounds on the Size of Walsh Coefficients}\label{sec:Negligibility}

We start by making a series of definitions that follow closely to Ref.~\cite{C-K72}. We assume throughout that the Walsh functions are Paley-ordered.

\begin{definition}\label{definition:rank}
Rank, degree, and negligibility
\end{definition}
For a natural number $m$ with binary expansion $m=m_nm_{n-1}...m_2m_1$, we define the rank, degree, and negligibility of $m$ as follows. The rank of $m$, denoted $r(m)$, is given by 
\begin{align}
r(m)=\sum_k m_k,
\end{align}
the degree of $m$, $d(m)$, is given by
\begin{align}
d(m)= \text{min}\{k:2^k>m\},
\end{align}
and the negligibility of $m$, denoted $p(m)$, is given by 
\begin{align}
p(m)=\left[\sum_{k=1}^nm_kk\right] + r(m).
\end{align}

Note that if the rank is large, then so are the degree and negligibility. However, large degree does not imply large rank, and also does not imply large negligibility. One of the key results that we will use is the following theorem that was stated and proved in~\cite{C-K72}. 
\begin{theorem}\label{thm:boundcoeff}
\end{theorem}
The size of the Walsh coefficients in Paley ordering for an analytic (smooth) function $f(t)$ depends on the rank and negligibility through either of the following inequalities:
\begin{align}
\left|\hat{f}_m\right| &\leq 2^{-p(m)} T^{r(m)}\text{max}_{t\in[0,T]}\left|f^{(r(m))}(t)\right|,\nonumber \\
\left|\hat{f}_m\right| &\leq 
2^{1-p(m)} T^{r(m)-1}V_0^T\left[f^{(r(m)-1)}\right].
\end{align}

\begin{figure}[t]
\begin{center}
\includegraphics[width=0.45\textwidth]{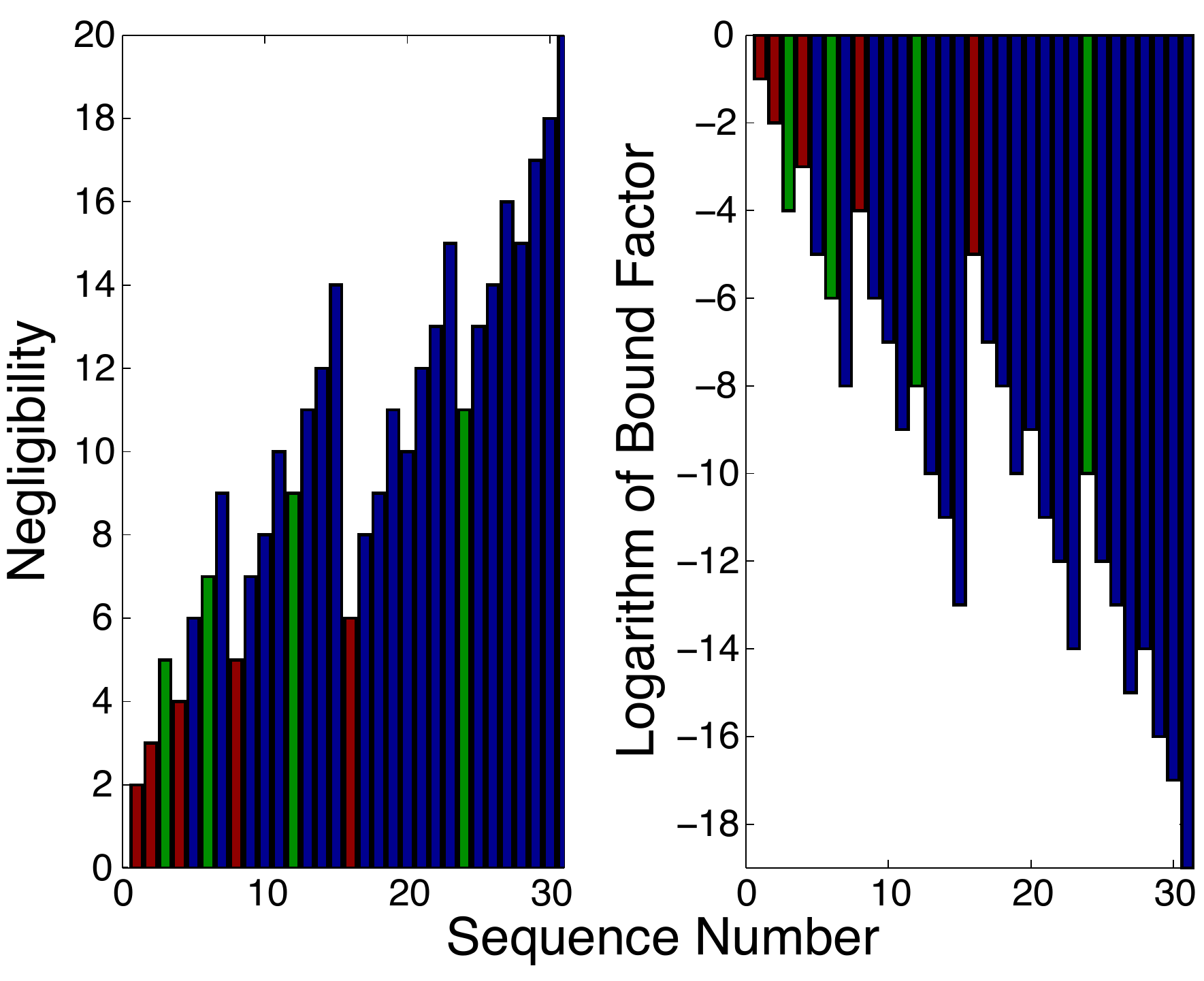}
\caption{\label{Fig:neg} Negligibility ($p(m)$) and logarithm of the bound factor ($1-p(m)$) for the first 32 Walsh sequences in Paley ordering. PDD are in red and CPMG in green.}
\end{center}
\end{figure}

We call $2^{1-p(m)}$ the bound factor of the $m$'th Walsh coefficient. Figure~\ref{Fig:neg} shows plots of the negligibility and logarithm of the bound factor for each of the first 32 Walsh functions. Clearly these two quantities are symmetric and, as $m$ grows larger, the bound factor decays to 0 in a non-monotonic fashion. Indeed, $2^{1-p(m)}$ decreases as $m$ goes from 0 to 3, then from 4 to 7, then again from 8 to 11, and so on. Increases are observed when passing from 3 to 4, 7 to 8, and so on. Moreover, there is an infinite subsequence of local maxima in $\{2^{1-p(m)}\}_{m=0}^{\infty}$, which is equivalent to an infinite subsequence of local minima in $\{p(m)\}_{m=0}^{\infty}$. We will make these features more explicit by first looking at the behavior of the negligibility in terms of the CPMG and PDD Walsh functions.

\subsection{Negligibility of CPMG and PDD Functions}\label{sec:NegligibilityCPMGPDD}

As previously described, the CPMG and PDD subsets of the Walsh basis play an important role in quantum information theory. From Subsection~\ref{sec:CPMGandPDD} we know that there are only $n-1$ CPMG and $n$ PDD functions in the first $2^n$ Walsh functions. Hence, the number of CPMG and PDD functions only grows logarithmically within the set of all Walsh functions and so are far from constituting a basis of piecewise constant functions. Because of this, a reconstruction based solely on estimating the CPMG and PDD coefficients can be highly inaccurate, especially when there is no simple symmetry in the function to be estimated (such as in a sine or cosine function). This is discussed in more detail with numerical examples in Subsection~\ref{sec:CPMGandPDDComparisons}. For now we use the negligibility function, $p(m)$, introduced in Def.~\ref{definition:rank}, to quantify how much information is generically contained in the CPMG and PDD coefficients. 
For simplicity, the indices for the PDD Walsh functions are denoted 
\begin{align}
\{g_j\}_{j=1}^{\infty} &= \{2^{j-1}\}_{j=1}^\infty =\{1,2,4,8,16,...\}
\end{align}
and the set of indices for the CPMG Walsh functions by 
\begin{align}
\{h_j\}_{j=1}^{\infty} &= \left\{\frac{3}{2}2^j\right\}_{j=1}^\infty = \{3,6,12,24,...\}.
\end{align}

It is straightforward to obtain an analytic expression for the negligibility of the the PDD and CPMG coefficients, $p(g_j)$ and $p(h_j)$. First, we note that the binary sequence which corresponds to the $j$'th PDD Walsh function, $g_j$, is zero everywhere except in the $j$'th entry. This gives 
\begin{equation}
p(g_j)=j+1\label{eq:negPDD}
\end{equation}
and so the bound factor for PDD sequences is given by
\begin{equation}
2^{1-p(g_j)}=2^{-j}.
\end{equation}
Next, the binary sequence which corresponds to the $j$'th CPMG function, $h_j$, is zero everywhere except in the $j$'th and $(j-1)$'th entries. Hence, for each $j\geq 1$
 \begin{equation}
 p(h_j)=(j-1)+j + 2=2j+1,\label{eq:negCPMG}
 \end{equation}
 and so the bound factor for CPMG sequences is given by
 \begin{equation}
 2^{1-p(h_j)}=4^{-j}.
 \end{equation}
 
Fig.~\ref{Fig:neg} provides  intuition for the behavior of the PDD and CPMG functions. In particular, the PDD and CPMG functions correspond to a (strict) infinite subset of the complete set of local maxima of $\left\{2^{1-p(j)}\right\}_{j=0}^{\infty}$. The subset is strict because $j=20$ and $j=28$ are also local maxima. Equivalently, the PDD and CPMG functions correspond to \emph{local minima} in the sequence $\{p(j)\}_{j=0}^{\infty}$. Thus, these sequences contain a large amount of information relative to nearby sequences. One may also guess from Fig.~\ref{Fig:neg} that the rate of divergence of $\{p(j)\}_{j=0}^\infty$ to $\infty$ is determined by the rate at which the negligibility of the PDD sequence diverges. This is indeed the case which gives a simple interpretation of the significance of the CPMG and PDD sequences within the set of all Walsh coefficients. We formally state and prove these results in Theorem~\ref{thm:neg}.

\begin{theorem}\label{thm:neg}
\end{theorem}
The following are true:
\begin{enumerate}
\item A local minimum in the negligibility sequence $\{p(j)\}_{j=1}^{\infty}$ occurs at $k$ if and only if a local minimum in the rank sequence $\{r(j)\}_{j=1}^{\infty}$ occurs at $k$.
\item The indices that correspond to PDD and CPMG sequences, $\{g_j\}$ and $\{h_j\}$, are local minima in the negligibility and rank sequences.
\item The rate of growth of any subsequence $\{p(j_k)\}_{k=0}^{\infty}$ of $\{p(j)\}_{j=0}^{\infty}$
is minimal on the set of PDD coefficients. Equivalently, the slowest possible rate of divergence of the negligibility over all possible subsequences is given by the PDD subsequence. 
\end{enumerate}

\begin{proof}

\bigskip
Proof of 1. First, suppose a local minimum in $\{p(j)\}_{j=0}^\infty$ occurs at $k$. We look at the binary sequences of $k-1$, $k$, and $k+1$. From the expression of the negligibility, without loss of generality, we can truncate the sequences starting from the leftmost digit on which $k-1$ and $k$ differ. Thus we can assume $k$ is of the form $1000....000$, $k-1$ is of the form $0111....1111$, and $k+1$ is of the form $1000....0001$ where the sequences all have the same length. Clearly $k$ has a local minimum in the rank. Conversely, suppose the rank takes a local minimum at $k$. Again, without loss of generality, we can assume $k$, $k-1$, and $k+1$ are of the above form and so $k$ is a local minimum in the negligibility.

\bigskip

Proof of 2. We have for any $k$
\begin{align}
p(g_k)&=k+1=\log(g_k)+2,\\
p(h_k)&=2k+1.
\end{align}
It is straightforward to verify that
\begin{align}
 p(g_k-1)&=\frac{k^2-1}{2},\nonumber \\
 p(g_k+1)&=k+3,\nonumber \\
p(h_k-1)&=\frac{k(k+1)}{2},\nonumber \\ 
p(h_k+1)&=2k+3. 
\end{align}
Thus, PDD and CPMG functions correspond to local minima in the negligibility. From Statement 1 of the theorem, they are also local minima in the rank.

\bigskip

Proof of 3. First, from Eq.~(\ref{eq:negPDD}), we have that $\{p(g_j)\}_{j=1}^\infty$ is a monotonically increasing subsequence of $\{p(k)\}_{k=0}^\infty$. We show that for any $k$ not corresponding to a PDD function, there exists $j$ such that $g_j > k$ and $p(g_j) < p(k)$. Thus, \emph{any} subsequence of $\{p(k)\}_{k=0}^{\infty}$ (not necessarily monotonic) is bounded below by the monotonically increasing negligibility of the PDD subsequence, $\{p(g_j)\}_{j=1}^{\infty}$. Hence, the rate of growth of any subsequence $\{p(j_k)\}_{k=0}^{\infty}$ of $\{p(j)\}_{j=0}^{\infty}$ is greater than or equal to the rate of growth of $\{p(g_k)\}_{k=1}^{\infty}$.

So, let us show that for any non-PDD index $k$, there exists $j$ such that $g_j > k$ and $p(g_j)< p(k)$. Since $k$ is non-PDD, the binary expansion of $k$ has at least rank 2. Suppose the binary expansion of $k$ is 
\begin{align}
....k_{i+1}k_i....k_2k_1
\end{align}
where $k_i$ is the leftmost entry that is equal to 1, i.e., $k_j=0$ for every $j \geq i+1$. The minimal value of $p(k)$ is equal to $(i+1)+2 = i+3$. The next largest index that corresponds to a PDD function corresponds to the binary string that is zero everywhere except at $k_{i+1}$. This has negligibility $i+2$. Since $i+2 < i+3$, we are done. 
\end{proof}

We can now state two corollaries of the above theorem that are easy to verify. First, we recall that for each $d$ there are $2^{d-1}$ Walsh functions that have degree $d$. Hence, there are $2^{d}$ Walsh functions of degree less than or equal to $d$. 

\begin{corollary}\label{thm:cor1}
\end{corollary}
Suppose at each step of moving to larger degree $d$, one is allowed to measure one and only one new Walsh coefficient to obtain maximum information about the function of interest. Then, at each step, the Walsh coefficient corresponding to the PDD Walsh function should be chosen as it minimizes the negligibility on the set of new Walsh functions.

\bigskip

Corollary~\ref{thm:cor1} implies that at each degree $d$, the PDD Walsh functions have the most significant information about a function in terms of the Walsh representation. Interestingly, if one extends the above corollary to being allowed to estimate any two coefficients, it is easy to see that the CPMG coefficient is \emph{not} the second best to choose in terms of minimizing negligibility. Indeed, $h_j$ (which indexes the $j$'th CPMG Walsh function) has degree equal to $j+1$. Moreover, $h_j$ is equal to 1 at each of the $j$ and $j+1$ entries, and is 0 elsewhere. Thus, any rank-2 binary sequence with a value of 1 in its $j$ and $k$'th entries where $k<j$ has smaller negligibility than $h_j$ (there are $j-1$ such sequences). This indicates that Walsh coefficients which do not correspond to PDD and CPMG sequences can be very significant in the Walsh reconstruction of a function. We collect these observations in Corollary~\ref{thm:cor2}.

\begin{corollary}\label{thm:cor2}
\end{corollary}
In each step of moving from degree $d-1$ to $d$, while the $d$'th PDD function minimizes the negligibility on the new set of $2^{d-1}$ coefficients, the $d-1$'th CPMG function does not provide the next largest negligibility on the set of new coefficients. In fact, since the new degree is equal to $d$, there are $d-2$ Walsh functions that have smaller negligibility than the new $d-1$'th CPMG function.

\bigskip

The results presented in this Subsection have provided us with an understanding of the behavior of the negligibility function on the CPMG and PDD Walsh functions. An even more general picture of the negligibility function is presented next in Subsec.~\ref{sec:Mapping}. 

\subsection{Minima in the Negligibility Function}\label{sec:Mapping}

We can use similar ideas from Theorem~\ref{thm:neg} and Corollaries~\ref{thm:cor1} and~\ref{thm:cor2} to provide a more complete picture of the negligibility function. The motivation for this is to understand what bounds can be placed on different Walsh coefficients of a function, which provides the basis for the data compression scheme based on threshold sampling in Sec.~\ref{sec:Sampling}. We know that, at each degree $d$, the minimal negligibility occurs at the $d$'th PDD function, which corresponds to the binary vector that is only non-zero in the $d$'th entry. In addition, there are $d-1$ Walsh functions with rank equal to 2, and the negligibility of these functions increases up to the maximum value at the $d-1$'th CPMG sequence (which we recall is only non-zero in the $d$'th and $d-1$'th entries of its associated binary vector). 

There is a much more structured behavior of the negligibility function on the entire set of Walsh functions. In particular, it is easy to see that large rank Walsh functions can have smaller negligibility than small rank Walsh functions. Here we analyze this finer structure of the negligibility and start with the following theorem.

\begin{theorem}\label{thm:neg2}
\end{theorem}
The following statements are true:
\begin{enumerate}
\item Let $\{k_j\}_{j=0}^\infty$ denote the subsequence of the natural numbers with $k_j=4j$. The local minima in the negligibility function occur exactly at each $k_j$. In addition, the sequence of local minima $\{k_j\}_{j=0}^\infty$ itself has local minima at the indices $j_l=4l$ for $l\geq 0$. Thus every fourth natural number corresponds to a minimum in the negligibility and every sixteenth natural number corresponds to a minimum in $\{k_j\}_{j=0}^\infty$. (one may guess this behavior from Fig.~\ref{Fig:neg}).
\item When $d\geq 3$, there are $2^{d-3}$ minima in the negligibility contained within the set of degree $d$ natural numbers. Hence the set of natural numbers with degree less than or equal to $d$ contains $2^{d-2}$ minima in the negligibility. 

\smallskip

When $d\geq 5$, there are $2^{d-5}$ minima in the minima of the negligibility that are contained within the set of degree $d$ natural numbers. Hence, the set of natural numbers with degree less than or equal to $d$ contains $2^{d-4}$ such minima.


\end{enumerate}

\begin{proof}

Proof of 1. Let $k_j=4j$ for $j\geq 0$. Clearly $k_0=0$ is a local minima in the negligibility. If $j\geq 1$, the natural number $4j-1$ has binary representation whose first two (rightmost) digits are $11$. Let the first digit that is equal to 0 in this binary representation appear in entry $q$. Then $k_j=4j$ has first non-zero entry at $q$, that is, it has degree equal to $q$. The difference in negligibility between $4j$ and $4j-1$ is
\begin{align}
p(4j)-p(4j-1) &= (1+q)-\left(q-1 + \sum_{i=1}^{q-1}i\right),\nonumber \\
&= 2-\frac{q(q-1)}{2}.
\end{align}
which is less than zero whenever $q\geq 3$ (i.e., $j\geq 1$). Next, the natural number $4j+1$ clearly has negligibility strictly greater than that of $4j$. Thus, for $j\geq 1$,
\begin{align}
p(4j-1)\geq p(4j)\leq p(4j+1)
\end{align}
which proves the first part of Statement 1. The second part of Statement 1 follows in exactly the same manner however now the binary representation of the natural numbers with indices $k_{j_l}-1$ have the value ``1" in their first four entries. 

%

\bigskip

Proof of 2. This follows from a simple counting argument. The local minima in negligibility occur at every fourth binary sequence and there are $2^{d-1}$ different sequences with degree $d$. Therefore there are $(2^{d-1})/4 = 2^{d-3}$ minima in negligibility. Similarly, the minima in the negligibility minima occur at every 16'th sequence. Since there are $2^{d-1}$ different sequences of degree $d$, there are $(2^{d-1})/16 = 2^{d-5}$ local minima in the set of local minima of the negligibility.
\end{proof}

\begin{figure}[b]
\begin{center}
\includegraphics[width=0.4\textwidth]{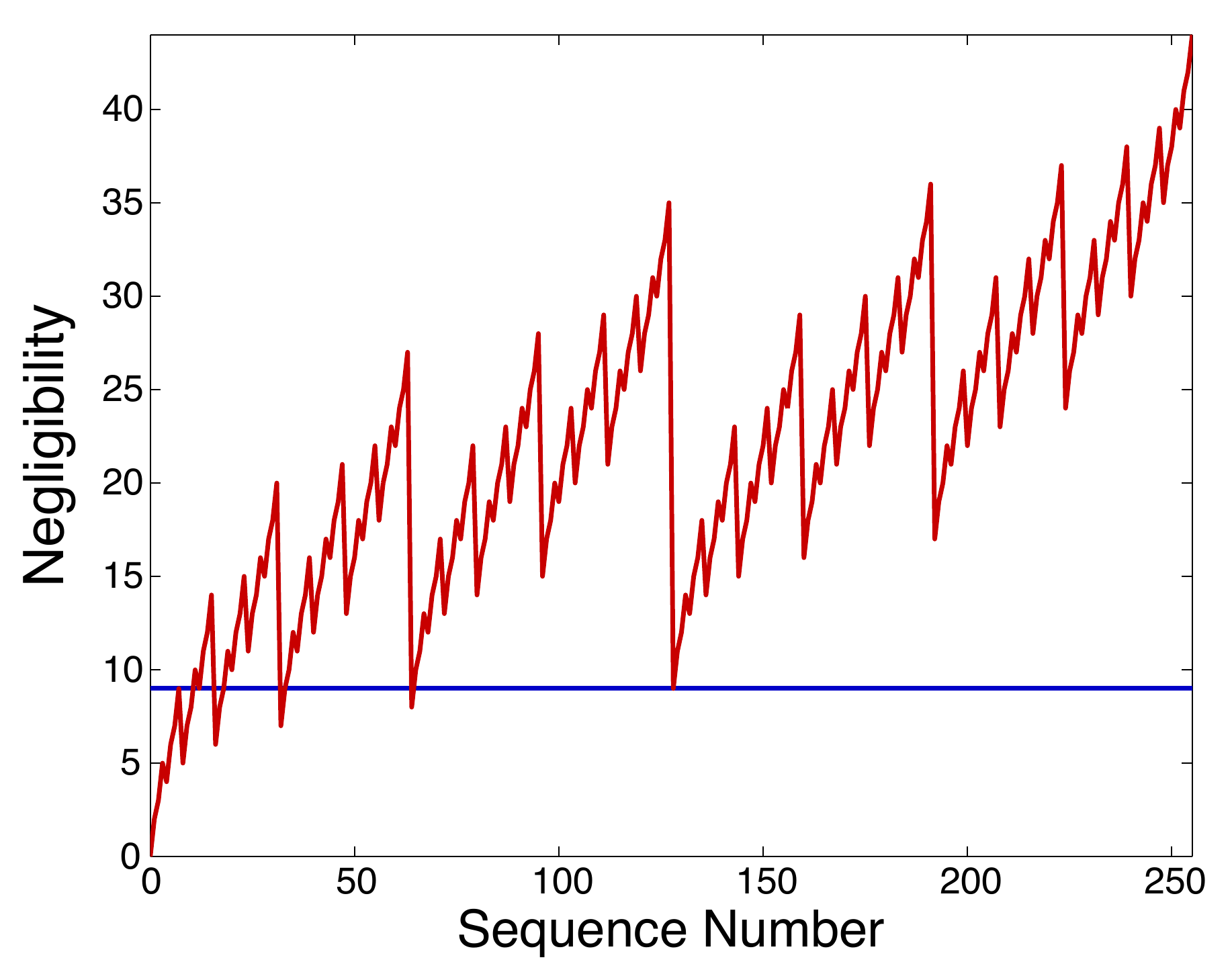}
\caption{\label{Fig:negconst} Negligibility for first $2^8$ natural numbers (red) compared with a constant negligibility of 9 (blue). Clearly most natural numbers have large negligibility and can potentially be ignored.}
\end{center}
\end{figure}

 To make the results of Theorem~\ref{thm:neg2} more transparent, a plot of the negligibility for all Walsh functions up to order 8 is given in Fig.~\ref{Fig:negconst}. The $m$'th Walsh function $\Wal_m$ such that $m$ has a binary representation of degree $d(m)=8$ starts at sequence number 128. There are $2^5=32$ minima in negligibility and $2^3=8$ minima in the minima over all degree 8 Walsh functions. 

We now introduce the notion of the ``contrast" of a natural number, which is of independent interest for understanding the role of CPMG and PDD Walsh functions within the entire Walsh basis. The contrast is a measure of how much the negligibility changes from $j-1$ to $j$.

\begin{definition} Contrast of a Natural Number
\end{definition}
The ``contrast" of a natural number j $\in \{1,2,3,...\}$, denoted $c(j)$, is defined as
\begin{align}
c(j)=p(j-1)-p(j).
\end{align}
\noindent This leads to the following simple corollary of Theorem~\ref{thm:neg2}.

\begin{corollary}\label{thm:cor3}
\end{corollary}

At each degree $d$, the two natural numbers with maximal contrast are those that correspond to the $d$'th PDD and CPMG Walsh functions.

\bigskip

Corollary~\ref{thm:cor3} states that if one is interested in maximizing the contrast, then PDD and CPMG functions are optimal.
The results of this section, and particularly, Theorems~\ref{thm:neg} and~\ref{thm:neg2}, provide insight into data compression methods based on the negligibility function that we formulate in the next section.

\subsection{Methods of Data Compression}\label{sec:Compression}

In this section, we discuss possible methods for data compression using Walsh representations. The motivation for understanding such methods is that the reconstruction of time-varying fields is often constrained by finite resources. Therefore, it is important to allocate the available resources for estimating the coefficients that provide the most information about the field. If the time-varying field can be accurately reconstructed using only a small number of coefficients, then the resources should be allocated for estimating only these coefficients. In what follows, we analyze three different methods for data compression; a CPMG/PDD based method, threshold sampling of coefficients, and the sub-degree method.



\subsubsection{CPMG and PDD Method}\label{sec:CPMGandPDDComparisons}
\begin{figure}[t]
\begin{center}
\includegraphics[width=0.45\textwidth]{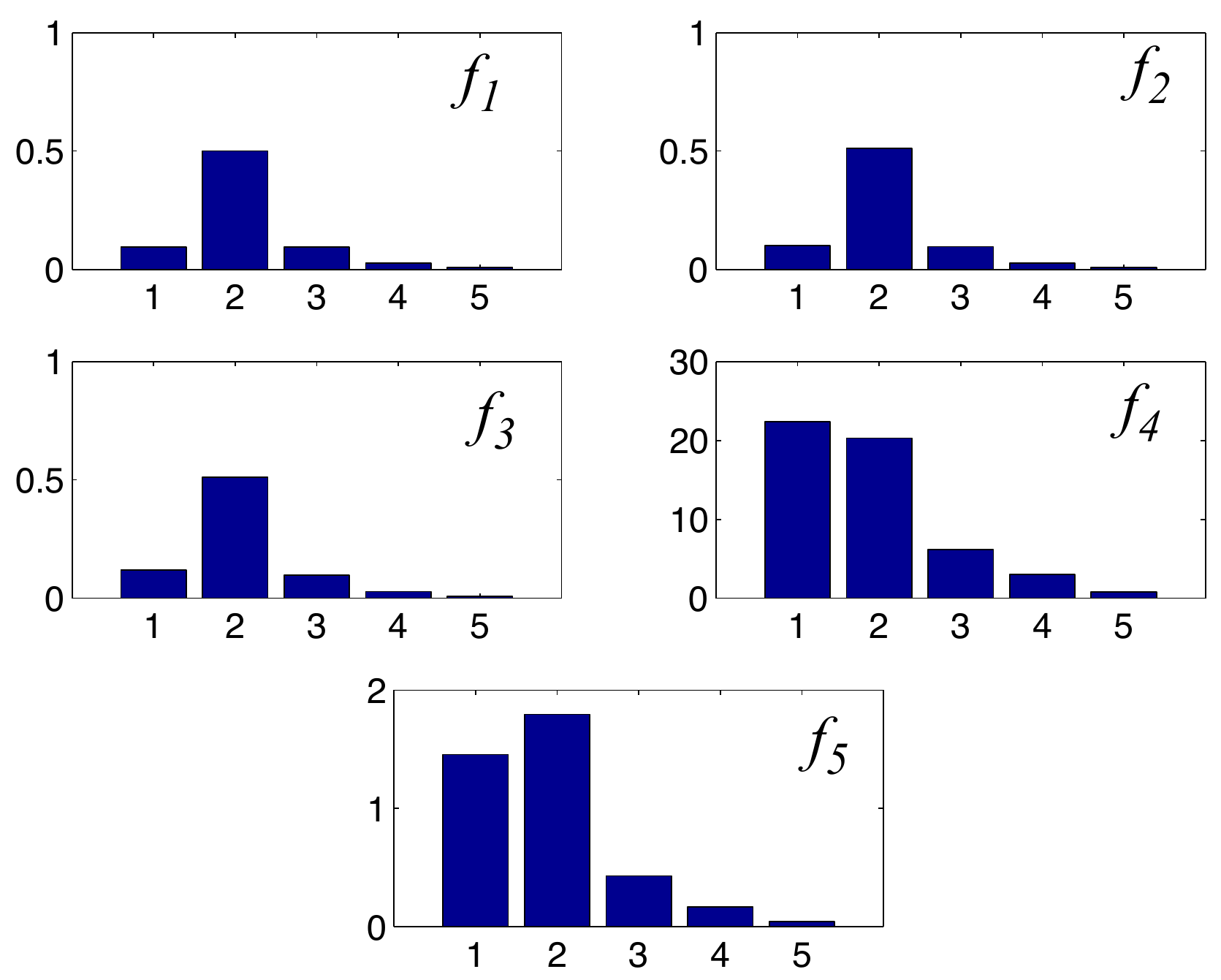}
\caption{\label{Fig:MSQE} Bar plot of MSQE for the five different functions $f_1$ through $f_5$.}
\end{center}
\end{figure}

As described in Section~\ref{sec:CPMGandPDD}, the CPMG and PDD functions are used for a variety of tasks in quantum information theory. From Sec.~\ref{sec:NegligibilityCPMGPDD} we know that these Walsh functions tend to have small negligibility and so can contain a large amount of information in the reconstruction of a function. Here, we analyze how well reconstructions based solely on CPMG and PDD Walsh functions perform relative to the Walsh reconstruction method with the coefficients sampled up to a finite order. The set of examples we looked at were
\begin{enumerate}
\item $f_1(t)=\cos(2\pi t)$,
\item $f_2(t)=\cos((2\pi + \epsilon)t)$, $\epsilon = 0.2$,
\item $f_3(t)=\cos((2\pi + \epsilon)t)$, $\epsilon = 0.5$,
\item $f_4(t)=2+3\cos(2\pi t)+4\cos(4\pi t)+6\sin(2\pi t)+2\sin(4\pi t)$,
\item $f_5(t)= \frac{1}{\sigma\sqrt{2\pi}}\exp\left(\frac{-(t-\mu)^2}{(2\sigma^2)}\right)$, $\mu = 0.3$, $\sigma = 0.1$,
\end{enumerate}
and the reconstruction methods we analyzed were
\begin{enumerate}
\item 8'th CPMG Partial Sum, 
\item 8'th PDD Partial Sum, 
\item 8'th CPMG and 8'th PDD Partial Sums (Walsh Coefficients and Walsh Functions),
\item 8'th Walsh Partial Sum (3rd order Walsh Reconstruction),
\item 16'th Walsh Partial Sum (4'th order Walsh Reconstruction).
\end{enumerate}
Note that we also included $w_0$ in the CPMG and PDD partial sums. We compared the reconstruction function $f_\text{rec}(t)$ with the original function $f(t)$ via the mean-squared error (MSQE) between $f_\text{rec}(t)$ and $f(t)$,
\begin{align}
\text{MSQE}(f_\text{rec}-f)&= \frac{1}{T}\int_0^T\left(f(t)-f_\text{rec}(t)\right)^2dt.
\end{align}
The results are shown in Fig.~\ref{Fig:MSQE}.
\bigskip




The first three examples correspond to a perfect cosine function and cosine functions with phase errors of $0.2$ and $0.5$. The MSQE for the 3rd and 4'th order Walsh reconstructions (methods 4 and 5) are relatively small and independent of the phase error. For instance, the MSQE for method 4 is consistently around $0.025-0.027$, which verifies that Walsh reconstructions are robust against phase errors. This is not true when one looks at the CPMG or PDD reconstructions since the MSQE can be large in these cases. It is important to note that the full Walsh reconstructions (Methods 4 and 5) outperform the CPMG and PDD reconstructions for all of $f_1$ through $f_5$. This is a direct indication that a large amount of information can be obtained with Walsh sequences that are not associated with conventional decoupling sequences.

Before moving on to the next data compression method, let us note some important parameters in implementing the different types of reconstructions. The methods that utilize an 8'th CPMG partial sum has $2^8-2=254$ zero-crossings, the methods that utilize an 8'th PDD partial sum has $2^9-(8+2)=502$ zero-crossings, and the methods that utilize both has $756$ zero-crossings. On the other hand, the 8'th and 16'th Walsh partial sums has $28$ and $120$ zero-crossings respectively. Because the number of crossings corresponds to the number of control $\pi$-pulses that need to be applied, the CPMG and PDD methods can be more susceptible to pulse errors.

\subsubsection{Threshold Sampling Method}\label{sec:Sampling}

Data compression via threshold sampling is based on the following situation. Suppose one is given some ``threshold" negligibility $p_0$, such that one would like to  compute only the Walsh coefficients whose associated negligibility is less than or equal to $p_0$. This can be of practical relevance because, ideally, one only wants to spend their resources on estimating Walsh coefficients that have significant contribution to the overall reconstruction. 

 Theorem~\ref{thm:boundcoeff} set bounds for the Walsh coefficients in terms of their negligibility $p(j)$  in Paley ordering. 
We assume that $f(t)$ is unknown \textit{a priori}, so that the only knowledge available for data compression is the negligibility function $p(j)$. 

We know from Theorem~\ref{thm:neg} that the natural number $2^{d-1}$ has degree equal to $d$ and has a negligibility of $d+1$. Moreover, the negligibility is strictly greater for \emph{all} larger natural numbers. Thus, if one was to perform a brute-force search of coefficients whose negligibility is no more than $p_0$, they would have to calculate the negligibility of
\begin{align}
\sum_{j=0}^{p_0-2}2^j + 1 = 2^{p_0-1} + 1
\end{align}
natural numbers, which scales exponentially in the threshold $p_0$. A simpler algorithm for finding the relevant coefficients to estimate is as follows. Suppose the threshold $p_0$ is given (and one has no knowledge of the different $\left|f^{(r(j))}(t)\right|$). For each degree $d$ up to and including $p_0-1$, do the following:

\medskip

\underline{Step 1}. In sequential order, compute and record $p(b_2)$ for the $d-1$ rank-2 sequences $b_2$ at degree level $d$. If $p(b_2) > p_0$ for some $b_2$, then stop and proceed to the next step.

\medskip

\underline{Step 2}. For each recorded rank-2 sequence $b_2$, compute and record $p(b_3)$ in sequential order for the $l(b_2)-1$ different rank-3 sequences associated to $b_2$. Here, $l(b_2)$ denotes the position of the lagging ``1" in the binary representation of $b_2$. If $p(b_3) > p_0$ for some $b_3$ then stop and move on to the next recorded rank-2 sequence. Once all rank-2 sequences from have been exhausted, proceed to the next step.

\medskip

\underline{Step 3}. Repeat Step 2 for increasing rank (up to the maximum value of $p_0-1$). If at some rank $r \leq p_0-1$, no new sequences are found then terminate the entire process.

\bigskip

Since $d\leq p_0-1$ this process terminates in finite time and it is not difficult to show that the time-complexity of such a procedure scales by at most $O\left(2^{\frac{p_0}{2}}\right)$. This is a slight improvement on the $O\left(2^{p_0}\right)$ scaling using a naive brute force search, however it is a loose bound, and we expect that the scaling is much smaller. We now look at some examples of the threshold method.

\bigskip


\underline{Example 1}

\bigskip

\begin{figure}[b]
\begin{center}
\includegraphics[width=0.4\textwidth]{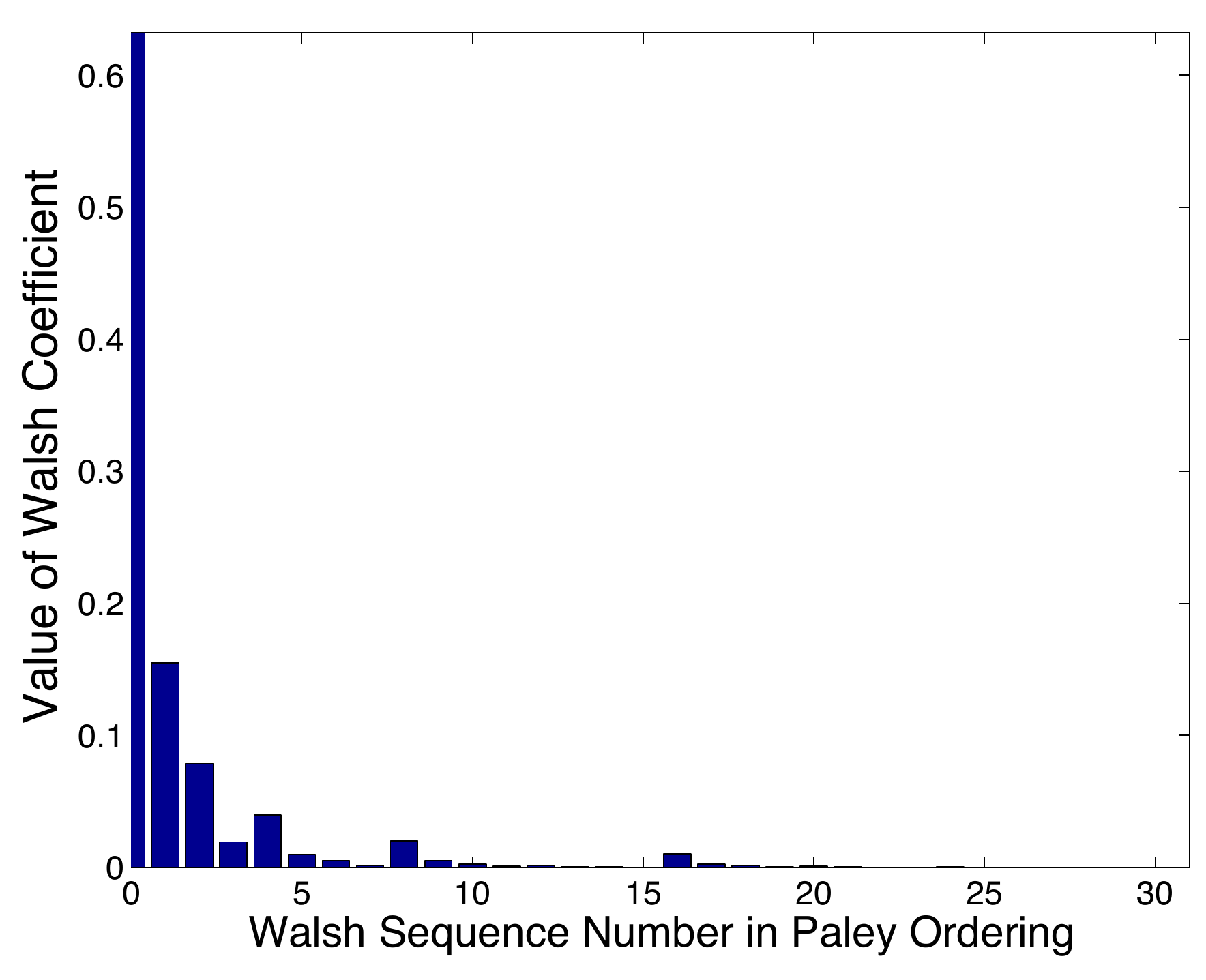}
\caption{\label{Fig:scatterexp} Log-plot of First 32 Walsh Coefficients for exp(-t).}
\end{center}
\end{figure}

\begin{figure}[t]
\begin{center}
\includegraphics[width=0.4\textwidth]{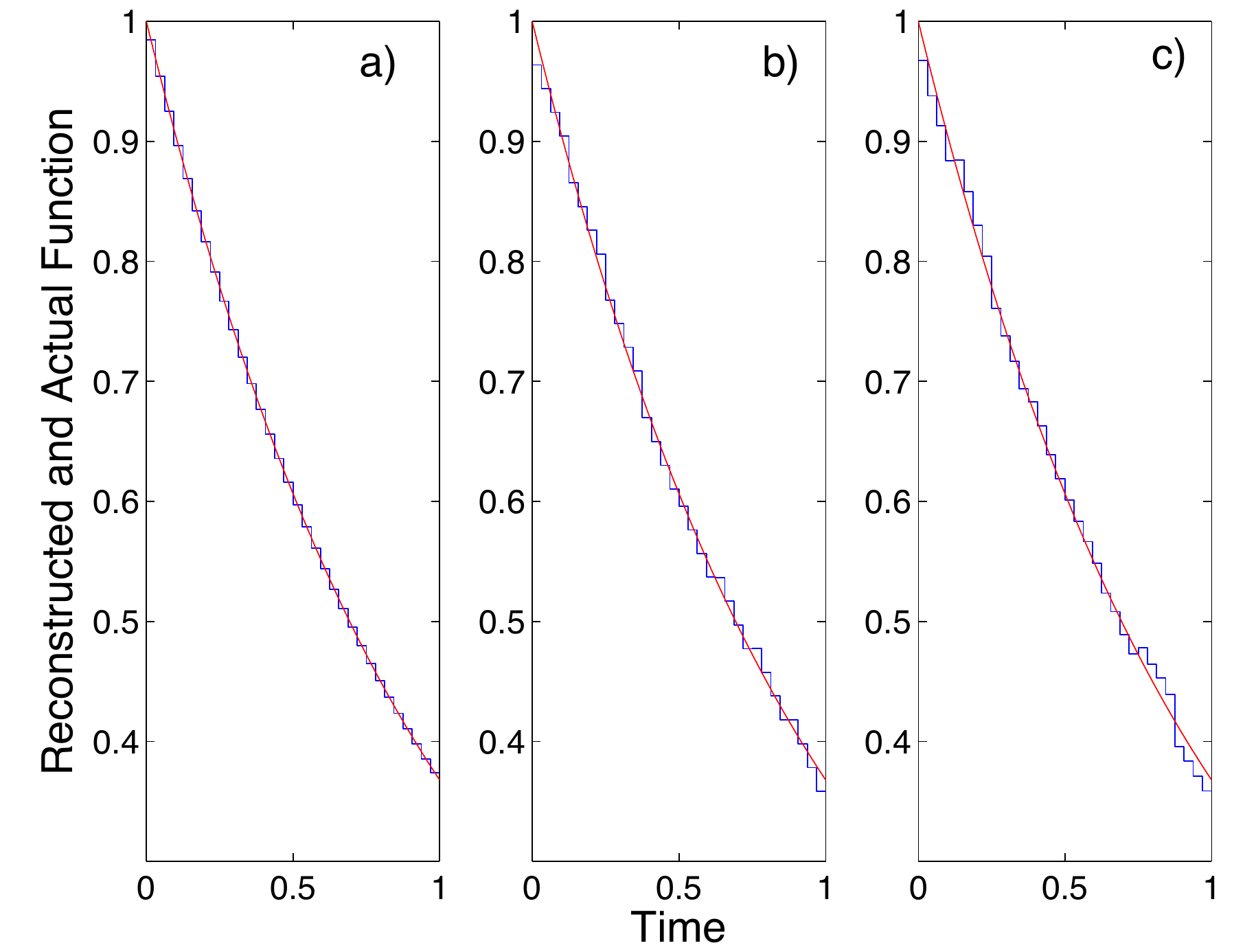}
\caption{\label{Fig:recexp} a) $5$'th order reconstruction (partial sum with 32 terms) and exact function exp(-t), b) Threshold-sampled reconstruction (threshold negligibility=6) and exact function, c) Sub-sampled reconstruction and exact function.}
\end{center}
\end{figure}

Let us first look at a simple example to clarify some of the concepts introduced in this Section. Suppose $f(t)=e^{-t}$ on $[0,1]$. Fig.~\ref{Fig:scatterexp} shows the magnitude of the first 32 Walsh coefficients of $f(t)$ in a log-plot. One can clearly see a pattern resembling that of the negligibility function in Fig.~\ref{Fig:neg}. Low rank coefficients have larger contribution to the reconstruction. In particular, PDD sequences (with sequence numbers corresponding to $\{1,2,4,8,...\}$) correspond to maxima in the value of the magnitude of the Walsh coefficients. There can however be counter-examples to this phenomenon, as we show in Example 2.
A plot of the complete $5$'th order reconstruction (partial sum with 32 terms), $f_{32}$, is given in Fig.~\ref{Fig:recexp}(a). As expected, $f_{32}(t)$ agrees with $f(t)$ very well, with MSQE($f_{32}-f$) =$3.518\times 10^{-5}$.

A plot of the threshold-sampled reconstruction for threshold negligibility $p_0=6$, denoted $f_{\text{th}}$, is given in Fig.~\ref{Fig:recexp}(b). The coefficients with negligibility less than or equal to 6 are: 0,1,2,3,4,5,8, and 16. Clearly, the reconstruction is still quite good and the mean square error between $f$ and $f_{\text{th}}$  is
$\text{MSQE}(f_{\text{th}}-f)=1.001\times 10^{-4}$.

\bigskip

\underline{Example 2}

\bigskip

\begin{figure}[b]
\begin{center}
\includegraphics[width=0.4\textwidth]{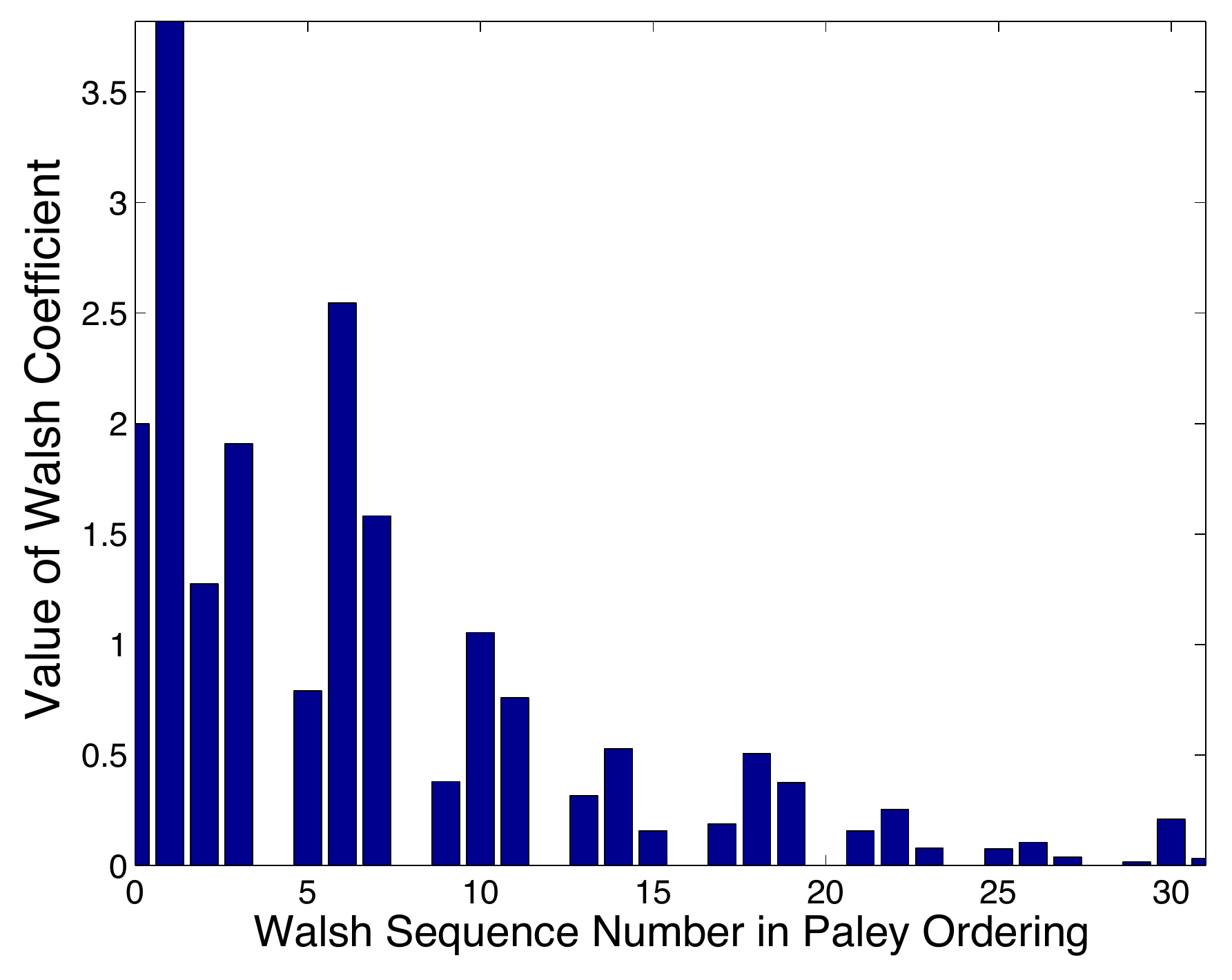}
\caption{\label{Fig:scatterpolychrom} First 32 Walsh Coefficients for $f_4(t)=2+3\cos(2\pi t)+4\cos(4\pi t)+6\sin(2\pi t)+2\sin(4\pi t)$.}
\end{center}
\end{figure}

\begin{figure}[b]
\begin{center}
\includegraphics[width=0.45\textwidth]{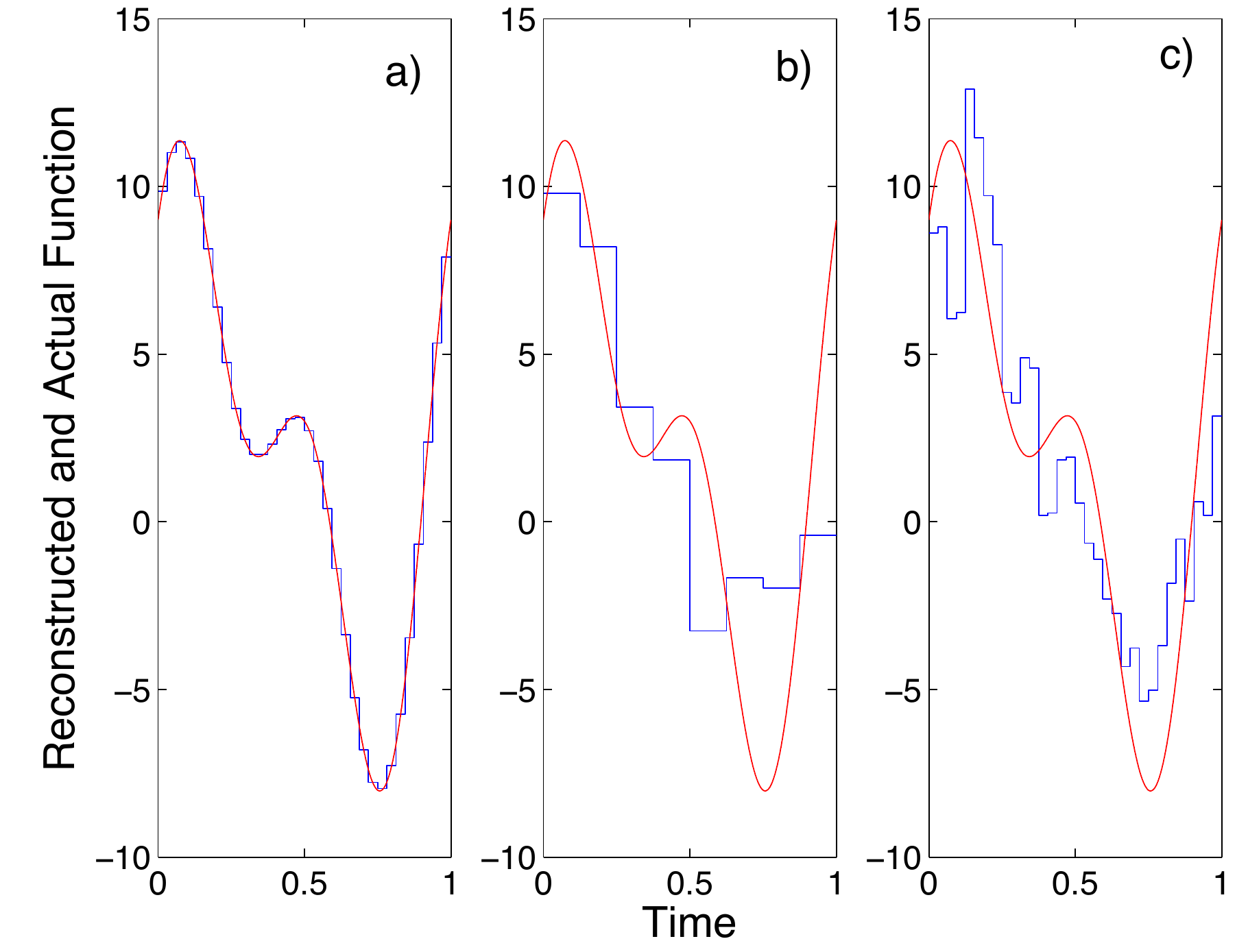}
\caption{\label{Fig:recpolychrom} a) Plot of the complete $5$'th order reconstruction (partial sum with 32 terms) and exact function  $f_4(t)=2+3\cos(2\pi t)+4\cos(4\pi t)+6\sin(2\pi t)+2\sin(4\pi t)$, b) Plot of the threshold-sampled reconstruction (threshold negligibility=6), c) Plot of the sub-sampled reconstruction.}
\end{center}
\end{figure}

Let us now look at an example involving a linear combination of simple trigonometric functions, $f(t)=f_4(t)$, discussed previously.
A log-plot of the absolute value of the first 32 Walsh coefficients for $f(t)$ is given in Fig.~\ref{Fig:scatterpolychrom}. 
Clearly, very few of these coefficients contribute to the reconstruction of $f(t)$. A 5'th order reconstruction of $f(t)$ using the first 32 coefficients is given in Fig.~\ref{Fig:recpolychrom}.
As expected, the 5'th order reconstruction agrees wih $f(t)$ very well and the mean-squared error between $f$ and $f_{32}$ in this case is
$
\text{MSQE}(f_{32}-f)=0.200.
$

We note that for Example 2, the non-negligible coefficients in Fig.~\ref{Fig:scatterpolychrom} do \emph{not} necessarily correspond to low rank coefficients. Hence, threshold sampling with $p_0=6$ produces a worse reconstruction of $f(t)$ than what we observed in Example 1. This is clear from Fig.~\ref{Fig:recpolychrom}c, with the mean-squared error $\text{MSQE}(f_{\text{th}}-f)=12.015$ which is quite large.
We conclude that threshold sampling will not always perform well. In fact, when the true function $f(t)$ is a simple trigonometric function (or linear combination of trigonometric functions), the reconstruction can fail badly. We anticipate that threshold sampling works well when the function has large support in the sequency (or frequency) domain. The third method for performing data compression, which we discuss below, deals with these cases much better, as seen in Fig.~\ref{Fig:recpolychrom}c.
\bigskip

\underline{Example 3}

\bigskip

\begin{figure}[t]
\begin{center}
\includegraphics[width=0.4\textwidth]{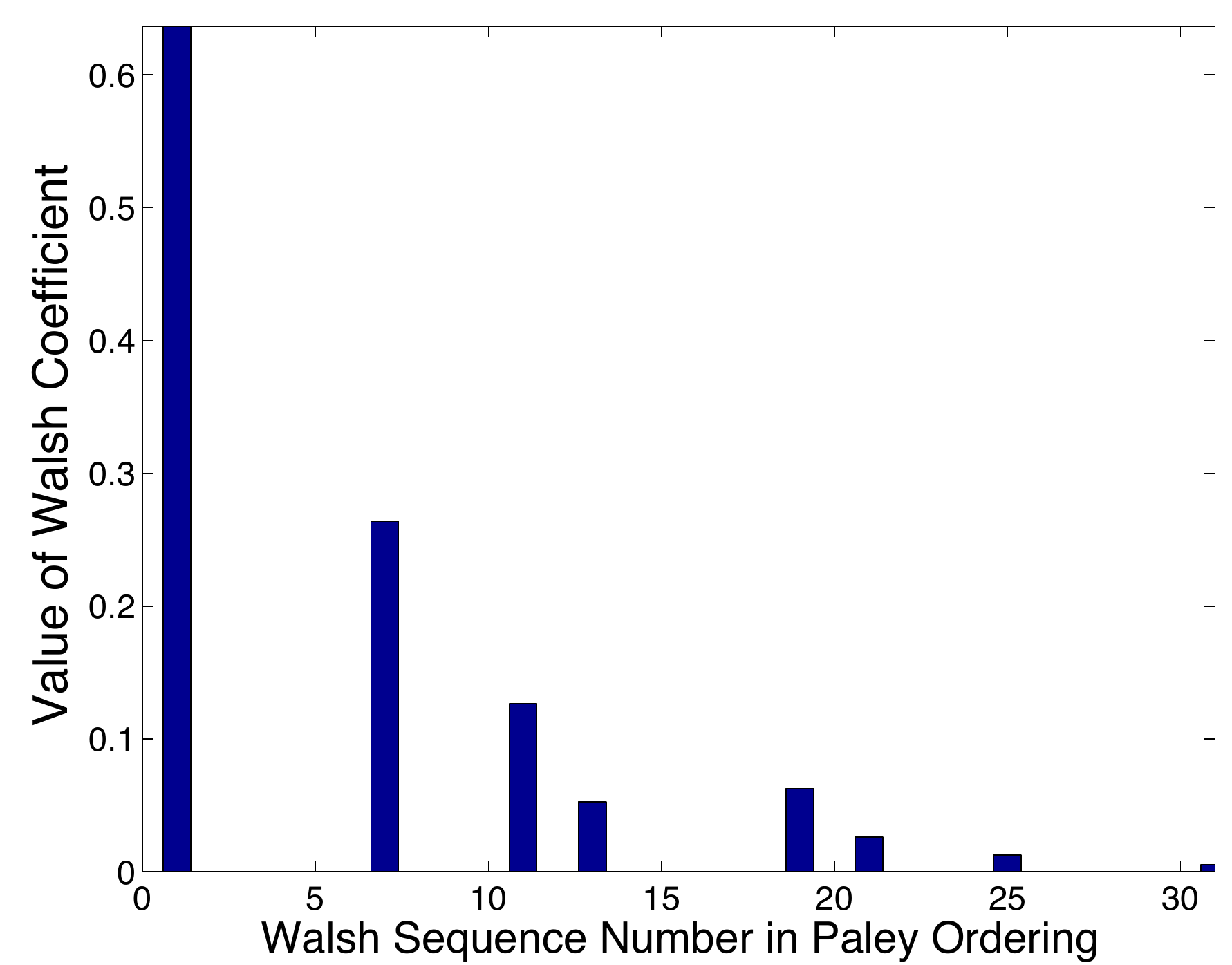}
\caption{\label{Fig:scattersin} First 32 Walsh coefficients for $\sin(2\pi t)$.}
\end{center}
\end{figure}

\begin{figure}[h!]
\begin{center}
\includegraphics[width=0.4\textwidth]{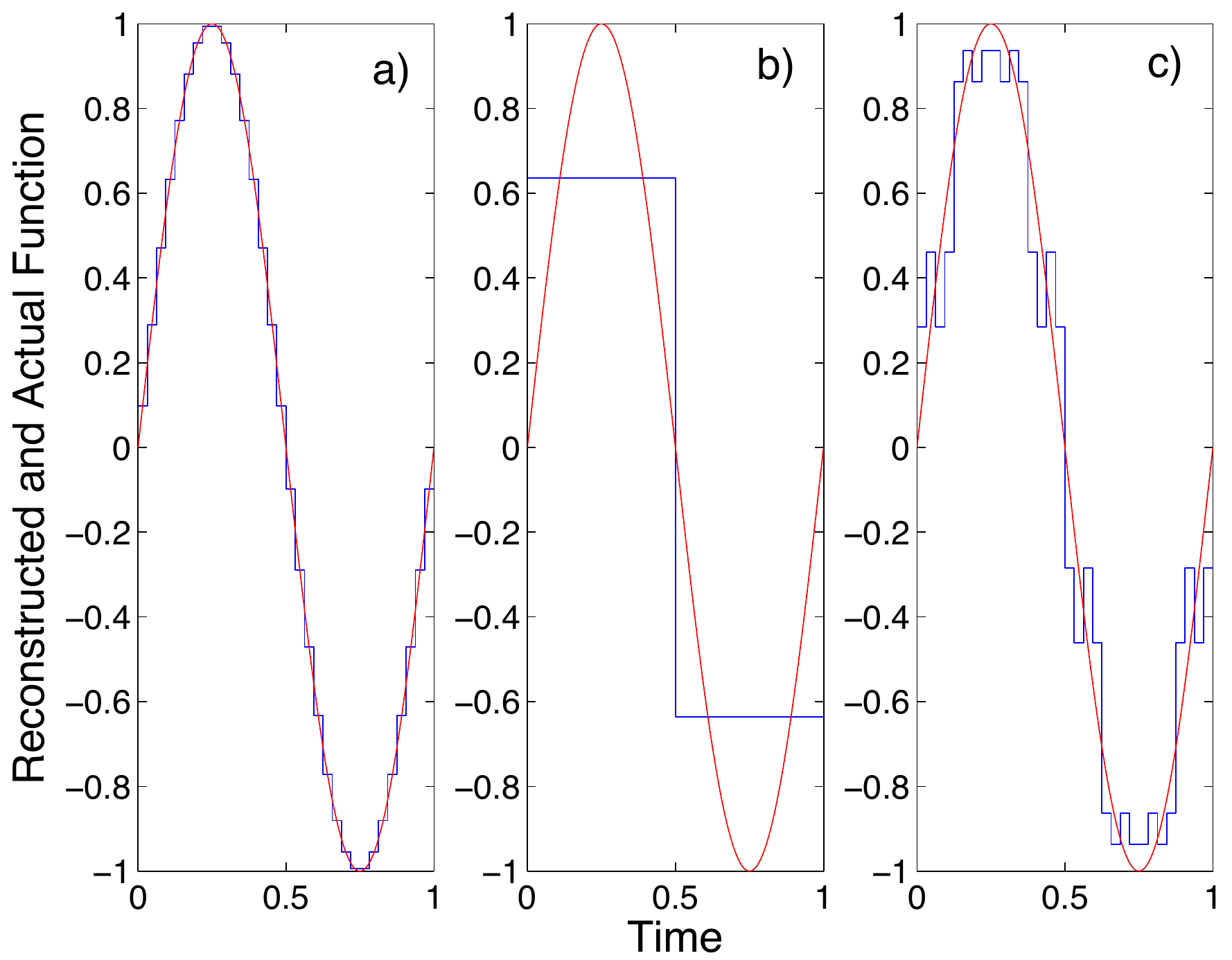}
\caption{\label{Fig:recsin} a) Plot of the complete $5$'th order reconstruction (partial sum with 32 terms) and exact function $\sin(2\pi t)$, b) Plot of the threshold-sampled reconstruction (threshold negligibility=6), c) Plot of the sub-sampled reconstruction.}
\end{center}
\end{figure}

The phenomenon of threshold sampling not performing well described in Example 2 can be made even more pronounced by using a single sinusoid $f(t)=\sin(2\pi t)$. A plot of the non-zero Walsh coefficients is given in Fig.~\ref{Fig:scattersin}.
The non-negligible coefficients occur at 1,7,11,13,19,21,25, and 31, which have negligibility values of 2,9,10,11,11,12, 13, and 20, respectively. Hence performing threshold sampling picks out only the first PDD coefficient and ignores the coefficients that contain the remaining information about $f(t)$ (see Fig.~\ref{Fig:recsin}). The mean-squared error of both $f_{32}$ and $f_\text{th}$ are
$
\text{MSQE}(f_{32}-f)=0.0016, 
\text{MSQE}(f_{\text{th}}-f)=0.0947.
$

\subsubsection{Sub-Degree Method}\label{sec:Sub-Degree}

We now look at a different method called the ``sub-degree method" for more efficient sub-sampling of Walsh coefficients, with the possibility for error estimates that were not possible via threshold sampling.

The sub-degree method, which has been introduced in Ref.~\cite{C-K75}, is an effective method for data compression that provides upper bounds on the error of the reconstruction. We define the sub-degree of a binary sequence of degree d to be the position of the second leading 1. Hence, the sub-degree of a degree $d$ sequence lies in $\{1,...,d-1\}$. Sequences with sub-degree $\gg$ 1 will have high rank and thus high negligibility. Suppose one wants to decide what coefficients to estimate at a particular degree $d$. Ref.~\cite{C-K75} has shown that ignoring all coefficients of sub-degree greater than $d^{\prime}$ (where $d^{\prime} \leq d-2$) leads to an error of at most 
\begin{align}
2^{-d^{\prime}-d-2}\left(1-2^{d^{\prime}-d+1}\right) T^2\sup_{t\in [0,1]} \left|f^{\prime \prime}(t) \right|.
\end{align}


Let us return to Examples 2 and 3, where threshold sampling did not perform well. In Example 2, we had $f(t)=f_4(t)$. For the first 32 Walsh coefficients, at each degree $d$ for $d=3,...,5$, we ignore all coefficients with sub-degrees greater than $d-2$. Thus, we only estimate 18 sequences and the resulting reconstruction, denoted $f_\text{ss}$, is given in Fig.~\ref{Fig:recpolychrom}. The mean-squared error for this method is given by
$
\text{MSQE}(f_{\text{ss}}-f) = 8.3339
$, 
which is still relatively large but smaller than $\text{MSQE}(f_{\text{th}}-f)=12.0145$.

In Example 3, $f(t)=\sin(2\pi t)$. Again, for the first 32 Walsh coefficients, at each degree $d$ for $d=3,...,5$, we ignore all coefficients with sub-degrees greater than $d-2$. The 1, 11, 19, and 21 Walsh coefficients are contained within the remaining 18 coefficients. The resulting reconstruction is shown in Fig.~\ref{Fig:recsin}, with the mean-squared error
\begin{equation}
\text{MSQE}(f_{\text{ss}}-f) = 0.0206,
\end{equation}
which is smaller than $\text{MSQE}(f_{\text{th}}-f)=0.0947$.

In conclusion, we have introduced three data compression methods based on (1) CPMG and PDD Walsh functions, (2) threshold sampling, and (3) the sub-degree method. These methods assumed no prior knowledge of the function $f(t)$; however, such knowledge can easily be incorporated into the three methods through the total variation of derivatives of $f(t)$. The advantages offered by each method vary according to the scenario at hand. When $f(t)$ is exactly equal to a sum of sine and cosine functions with no phase shift, a combination of CPMG or PDD based reconstruction will work well; however if phases are introduced, or the function is more complicated, this method will fail badly. Threshold sampling can work well when particular coefficients (or their negligibilities) are known to be extremely small. The sub-degree method is often more desirable than the other two methods, because it works well in most scenarios and provides an explicit bound on the reconstruction error. We note there also exist adaptive methods for data compression~\cite{C-K75} that may perform better than the non-adaptive methods that we have introduced here.

\section{Dynamical Decoupling With Walsh Sequences}\label{sec:Dynamical Decoupling}

In the previous section we looked at the structure of the negligibility function and provided the basis for simple data compression methods. In this section we discuss the dynamical decoupling~\cite{VKL} power of Walsh functions, a concept originally formulated in~\cite{HKV}. We use the negligibility function to unify the concepts of dynamical decoupling power and information content of Walsh functions. 

We first introduce the concepts required for analyzing the decoupling power of Walsh functions. Suppose the quantum system is affected by dephasing noise that can be modeled as a stationary stochastic process. Let $p$ be the sequence that contains the times at which control $\pi$-pulses are applied in the interval $[0,T]$. The coherence of the system at time $T$, under the decoupling sequence $p$, is given by~\cite{Uhrig08}
\begin{align}
W(\tau) &=\left|\overline{\langle \sigma_X\rangle (T)}\right| = e^{-\chi_p(T)}
\end{align}
where $\langle \sigma_X\rangle$ is the expectation value of the Pauli $X$ operator $\sigma_X$, the initial state of the system is $|+\rangle = \frac{1}{\sqrt{2}}(|0\rangle + |1\rangle)$, and $\chi_p(T)$ is given by
\begin{align}
\chi_p(T) = \frac{1}{\pi}\int_0^{\infty} \frac{S_{\beta}(\omega)}{\omega^2}F_p(\omega T)d\omega.\label{eq:powerspec}
\end{align}
$S_{\beta}(\omega)$ is the power spectral density (spectrum) of the noise and $F_p(\omega T)$ is the filter function in the frequency domain associated with the control sequence $p$. Determining the exact form of $\chi_p(T)$ requires a detailed knowledge of the power spectrum. However, one can directly compare the decoupling power of Walsh sequences via the behavior of their associated filter functions~\cite{Uhrig08,BDU}

In the frequency domain, the filter function associated with the Walsh function $m$ is denoted by $F_m (\omega T)$. For every $m$, $F_m (0)=0$, and the rate at which $F_m$ increases provides a characteristic of its performance in terms of blocking low frequency components of the noise. The main result of Ref.~\cite{HKV} was that Walsh functions of rank $r$ annihilate any polynomial of degree less than $r$, 

\begin{align}\label{eq:annihilate}
\int _0^1t^kw_m(t)dt = 0.
\end{align}
Thus,  
if the $m$'th Walsh function has rank $r$,
\begin{align}
F_m(\omega T) \sim (\omega T)^{2(r+1)}.
\end{align}

There are various implications of this result. First, high-rank Walsh filter functions are better high-pass filters than low rank Walsh filter functions~\cite{HKV}. More precisely, filter functions associated to high-rank Walsh functions have the sharpest roll-off at low frequencies and so annihilate a larger set of frequencies in the power spectrum of the noise as given by Eq.~(\ref{eq:powerspec}). It is important to note that, at any degree $d$, high-rank functions within the set $\{2^{d-1},...,2^d-1\}$ do not correspond to Walsh functions with the largest number of zero-crossings (sequency). Hence, rank and sequency should be kept as separate characteristics of Walsh functions. 
At each degree $d$, the Walsh functions with 1's in every entry of their binary representation correspond to the well-known concatenated dynamical decoupling (CDD) sequences, which are the $d$-fold concatenation of the single spin-echo sequence. 

While the rank gives an indication of the decoupling power of different Walsh functions, one would like a finer method to distinguish Walsh functions. For instance, even though (0,...,0,1,0,0,0) and (0,...,0,0,0,0,1) are both rank-1 sequences, the former will have a filter function with steeper roll-off and thus should be more optimal for decoupling. 
Here, we show that the \emph{negligibility} provides the ability to distinguish between the decoupling power of Walsh functions in a more detailed manner. More precisely, if two Walsh functions have the same rank $r$, then the one with the higher negligibility should have steeper roll-off. We make this intuition precise with the following result. Given the $m$'th Walsh function in Paley ordering, the exact second order expression for $F_m(\omega T)$ is given by,
$$\frac{\left(\omega T\right)^{2(r(m)+1)}}{4^{p(m)}}$$.
Hence, for two Walsh functions with the same rank $r$, the Walsh sequence with the larger negligibility will have steeper roll-off. To see why this is true, we have from Ref.~\cite{HKV} that for any reconstruction order $n$, if 
$
T_{\text{min}}:=\frac{T}{2^n},
$
then
\begin{align}
F_m(\omega T)&=4^{n+1}\sin^2\left(\frac{\omega\Tmin}{2}\right)\left[\Pi_{j|b_j=1}\sin^2\left(2^{n-j-1}\omega\Tmin\right)\right]
\\\nonumber
&\times\left[\Pi_{j|b_j=0}\cos^2\left(2^{n-j-1}\omega\Tmin\right)\right].
\end{align}
Keeping terms to 2nd order, and using the definition of both the rank and negligibility, we have
\begin{align}
F_m(\omega T)&\sim4^{n+1}\left[\frac{\omega^2\Tmin^2}{4}\right]\left[\Pi_{j|b_j=1}\left(4^{n-j-1}\omega^2\Tmin^2\right)\right] \nonumber \\
&= 4^n\left[\left(\omega\Tmin\right)^{2(r+1)}\right]\left[\Pi_{j|b_j=1}4^{n-j-1}\right] \nonumber \\
&= \left[\left(\omega T\right)^{2(r+1)}\right]\left[4^{-\sum_{j|b_j=1}(j+1)}\right] \nonumber \\
&=\frac{\left(\omega T\right)^{2(r(m)+1)}}{4^{p(m)}}.
\end{align}

In Sec.~\ref{sec:Data Compression} we obtained results regarding bounds on the magnitude of Walsh coefficients of a function $f(t)$ through the negligibility. This can generically be thought of as the information content in $w_j$ for reconstructing $f(t)$. From above, we can now also explicitly characterize how well $w_j$ performs as a dynamical decoupling sequence. Here, we summarize these results which provides a nice duality between information content and roll-off of the filter function.

\bigskip

Given the $j$'th Walsh function, we have that

\begin{enumerate}
\item if we want to reconstruct the smooth function $f:[0,T]\rightarrow \mathbbm{R}$ using the Walsh basis, then the contribution of $w_j$, given by the $j$'th coefficient $\hat{f}_j$, satisfies each of
\begin{align}
\left|\hat{f}_j\right| &\leq \frac{\text{max}_{t\in[0,T]}\left|f^{(r(j))}(t)\right|T^{r(j)}}{2^{p(j)}},\nonumber \\
\left|\hat{f}_j\right| &\leq  \frac{2V_0^T\left[f^{(r(j)-1)}\right]T^{r(j)-1}}{2^{p(j)}};\label{eq:duality1}
\end{align}
\item if we want to use $w_j$ as a dynamical decoupling sequence, then the low frequency roll-off of the filter function $F_j(\omega T)$ associated with $w_j$ is given by
\begin{align}
F_j(\omega T) = \frac{\left(\omega T\right)^{2(r(j)+1)}}{4^{p(j)}}.\label{eq:duality2}
\end{align}
\end{enumerate}
\noindent Hence, if $p(j)$ is small, $w_j$ is generally a poor dynamical decoupling sequence, however it can extract significant information about a function through the magnitude of its associated Walsh coefficients. Similarly, if $p(j)$ is large, then $w_j$ is generally a good dynamical decoupling sequence, but it only extracts little information about a function through the magnitude of its associated Walsh coefficient. 

Finally, we see from Eq.~(\ref{eq:sensitivityT2}) and Eq.'s~(\ref{eq:duality1}), (\ref{eq:duality2}) that there is a trade-off between information content and dynamical decoupling when choosing the Walsh sequence that offers the best sensitivity: larger negligibility implies longer coherence times, which improves the sensitivity; but at the same time, larger negligibility can imply smaller Walsh coefficients, which degrades the sensitivity.

\bigskip

%
%

\section{Conclusion}\label{sec:Conclusion}
We have provided a comprehensive discussion of the theoretical concepts that form the basis for the experimental reconstruction of magnetic field profiles using the Walsh basis. We have given a brief introduction to properties of the Walsh basis and a detailed analysis of many statistical concepts of the Walsh reconstruction protocol, including the Fisher information and sensitivity in estimating field amplitudes. We showed that the additive nature of the Walsh transform implies that errors in estimating Walsh coefficients propagate in a straightforward manner to the time-domain reconstruction. The sensitivity analysis and error propagation together provide a method for characterizing how close the reconstructed field is to the true field. The reconstructed field is a time-dependent Gaussian random variable whose variance depends on the sensitivity and error in estimating Walsh coefficients.

In practical scenarios, finite resources imply that one should spend these resources to estimate Walsh coefficients that have a significant contribution to the reconstruction of the magnetic field. We have provided a comprehensive discussion of data compression techniques and the negligibility function, which helps characterize the extent to which individual Walsh coefficients contribute to the reconstruction. We analyzed three methods for data compression: CPMG and PDD methods, threshold sampling methods, and the sub-degree method. Various examples were provided for each method and we found that each can be useful in different scenarios. Overall, the sub-degree method is desirable as it works well in most cases and gives explicit bounds on the reconstruction error. Lastly, we discussed the role of the Walsh functions in dynamical decoupling and provided new results using the negligibility to show that Walsh functions whose associated filter functions have steep roll-off tend to have low information content and Walsh functions whose associated filter functions have shallow roll-off tend to have large information content. This provides a competing effect in the sensitivity of estimating magnetic fields using Walsh coefficients.

Some directions of future research include whether there are advantages in adapting the presented Walsh reconstruction method to other digital orthonormal bases, such as the Haar basis~\cite{Haar10}. In addition, it would be interesting to analyze the possibility of extending the methods presented here to performing spectroscopy of stochastic fields and environments. Using the Walsh basis could eliminate some of the difficulties with Fourier-based spectroscopic techniques that arise from attempting to match discrete control sequences with smooth trigonometric functions. More generally, as we have highlighted in this paper, many properties of the Walsh functions make them well-suited for problems related to quantum parameter estimation, quantum control, and active quantum error-correction. It will be useful to continue to nvestigate their applicability in these areas as well as within the broader context of quantum information theory.


\begin{acknowledgements}
The authors thank Jonathan Welch for helpful discussions. This work was supported in part by the U.S. Army Research Office through a MURI grant No. W911NF-11-1-0400 and by DARPA (QuASAR program). A. C. acknowledges support from the Natural Sciences and Engineering Research Council of Canada.
\end{acknowledgements}

%
%
%
%

\end{document}